\documentclass[%
 reprint,
 nofootinbib,
 amsmath,amssymb,
 aps,
 prd,
]{revtex4-2}

\usepackage{graphicx}
\usepackage[caption=false]{subfig}

\usepackage{dcolumn}
\usepackage{bm}

\usepackage{afterpage}
\usepackage{floatpag}

\usepackage[colorlinks=true,allcolors=blue]{hyperref}
\usepackage{breakurl}

\begin{document}


\title{Detection of Gravitational Waves Using Bayesian Neural Networks}

\author{Yu-Chiung Lin}
\author{Jiun-Huei Proty Wu}%
 \email{jhpw@phys.ntu.edu.tw}
\affiliation{%
Department of Physics, Institute of Astrophysics, and Center for Theoretical Physics, National Taiwan University, Taipei 10617, Taiwan
}%


\begin{abstract} 
We propose a new model of Bayesian Neural Networks to not only detect the events of compact binary coalescence in the observational data of gravitational waves (GW) but also identify the full length of the event duration including the inspiral stage.
This is achieved by incorporating the Bayesian approach into the CLDNN classifier, which integrates together the Convolutional Neural Network (CNN) and the Long Short-Term Memory Recurrent Neural Network (LSTM).
Our model successfully detect all seven BBH events in the LIGO Livingston O2 data, with the periods of their GW waveforms correctly labeled.
The ability of a Bayesian approach for uncertainty estimation enables a newly defined `awareness' state for recognizing the possible presence of signals of unknown types,
which is otherwise rejected in a non-Bayesian model.
Such data chunks labeled with the awareness state can then be further investigated rather than overlooked.
Performance tests with 40,960 training samples against 512 chunks of 8-second real noise mixed with mock signals of various optimal signal-to-noise ratio $0 \leq \rho_\text{opt} \leq 18$ show that
our model recognizes 90\% of the events when $\rho_\text{opt} >7$ (100\% when $\rho_\text{opt} >8.5$) and successfully labels more than 95\% of the waveform periods when $\rho_\text{opt} >8$.
The latency between the arrival of peak signal and generating an alert with the associated waveform period labeled is only about 20 seconds for an unoptimized code on a moderate GPU-equipped personal computer.
This makes our model possible for nearly real-time detection and for forecasting the coalescence events when assisted with deeper training on a larger dataset using the state-of-art HPCs.
\end{abstract}

\maketitle

\section{Introduction}\label{intro}
Since the first detection of gravitational waves (GWs) on September 14th, 2015 \cite{PhysRevLett.116.061102}, the Advanced Laser Interferometer Gravitational Wave Observatory (aLIGO) \cite{LIGO2015}, later joined by the Advanced Virgo \cite{Acernese_2014} in 2017, has detected thirteen coalescence events for binary black holes \cite{PhysRevLett.116.061102,PhysRevLett.116.241103,PhysRevX.6.041015,PhysRevLett.118.221101,Abbott_2017,PhysRevLett.119.141101,LIGOScientific:2018mvr,Abbott:2020tfl,Abbott:2020khf,LIGOScientific:2020stg} and two for binary neutron stars \cite{TheLIGOScientific:2017qsa,Abbott:2020uma} during its O1, O2 and ongoing O3 observation runs. 
The triumph also comes with a remarkable milestone where the electromagnetic (EM) counterparts of the binary neutron star coalescence event GW170817 \cite{TheLIGOScientific:2017qsa} was discovered \cite{Goldstein_2017,Savchenko_2017,Abbott_2017_GW_GRB}. 
The success of LIGO and Virgo has opened a new era of multi-messenger astronomy, allowing for independent measurement of Hubble constant \cite{Soares-Santos:2019irc} and constraints on theoretical models such as cosmic strings \cite{Abbott:2017mem}. In addition, KAGRA \cite{Akutsu:2018axf}, a GW observatory in Japan, started observation in February 2020 and another GW detector located in India \cite{indigo} is also about to join the network in order to increase both the overall sensitivity and the precision in sky source locations. While more detectors join the network, the need for techniques of real-time detection has become more pressing not only for accurate determination of sky source locations but also for the counterpart observations such as those for EM signals \cite{Abbott2018}.
For example, the searches for EM counterparts such as the kilonovae and short-gamma-ray bursts \cite{Metzger_2012} can improve the accuracy of source parameter estimation while increasing the confidence for GW detections \cite{Chan:2015bma,Blackburn_2015}, and help break the modeling degeneracy of binary properties \cite{Chan:2015bma,Pankow_2017} so as to understand better the nature of binary systems and their host galaxies \cite{Coughlin:2018lta}.
However, the commonly used match-filtering techniques \cite{Owen:1998dk,pycbc,Usman_2016,Sachdev:2019vvd,PhysRevD.95.042001,Cannon:2011vi} are computationally expensive, making it a great challenge for real-time detection.

Recently the deep learning technique based on artificial neural networks \cite{Hertz:1991:ITN:104000} is considered as a promising alternative to the matched-filtering method. Various types of Deep Neural Networks (DNNs) such as the Convolutional Neural Network (CNN) \cite{6795724} and the Long Short-Term Memory Recurrent Neural Network (LSTM) \cite{doi:10.1162/neco.1997.9.8.1735, lstm,doi:10.1162/089976600300015015} have shown great potential in the framework of GW research, especially for real-time detection \cite{George:2016hay,George:2017pmj,Gebhard:2019ldz,Gabbard:2017lja,2019SCPMA..6269512F,Krastev:2019koe,2019arXiv191010525L}, parameter estimation \cite{George:2016hay,George:2017pmj,2019SCPMA..6269512F,Gabbard:2019rde}, glitch recognition \cite{Zevin:2016qwy,George:2017fbn,Llorens-Monteagudo:2018ubm,Coughlin:2019ref} and data denoising \cite{Shen:2019ohi,Wei:2019zlc,Shen:2017jkj,PhysRevResearch.2.033066}. However, despite their potential prospect, the DNN models usually suffer from overfitting and thus are difficult to be generalized for making reliable predictions in face of the data that are out of distribution. In addition, most DNN models are deterministic in a way that they offer only one rigid prediction for each given set of input. A deterministic model can hardly provide information about the uncertainty in its predictions. Therefore the problems of overfitting and lack of uncertainty estimation make the DNN models unreliable, sometimes giving overconfident predictions on out-of-distribution data \cite{nguyen2015deep}.

In order to deliver the uncertainty information, we need to convert the traditional deterministic model into a probabilistic model. In the field of deep learning, the Bayesian Neural Network (BNN) \cite{MacKay:1992:PBF:148147.148165, Hinton1995BayesianLF}, which updates its weights via Bayes' rule, is a potential choice for this purpose. The BNN technique is not only capable of  providing uncertainty estimation but also resistant to overfitting. In addition, it can be trained with a rather small dataset. The uncertainty estimation is particularly useful in face of the out-of-distribution data and could assist further training for data augmentation.
Although the BNN is more computationally expensive than the deterministic DNN, it is totally feasible when assisted with the recent-year advances in both the hardware and the approximation algorithms \cite{2015arXiv150602142G,Graves:2011:PVI:2986459.2986721,2015arXiv150505424B,2018arXiv180304386W,2013arXiv1312.6114K,2015arXiv150602557K}.

In this paper we use the Variational Inference (VI) approximation \cite{Graves:2011:PVI:2986459.2986721,2017arXiv170402798F,2015arXiv150505424B,2018arXiv180304386W} to construct a Convolutional, Long Short-Term Memory, Fully-Connected Deep Neural Network (CLDNN) model \cite{7178838}.
The incorporation of the sliding-window search scheme enables us to identify the time period of the GW waveform in a coalescence event. This feature distinguishes our model from other CNN models for GW detection in literature.
The combination of a Bayesian approach and the CLDNN is also a unique feature of our model.
Based on the model prediction and its estimated uncertainty, our Bayesian model can rapidly flag each of the sequential time windows with a trigger state, a noise state, or a state that needs further attention. The estimated uncertainty can also serve as a reference for the significance level of a prediction. 
Our model is not to replace the matched-filtering search or other models of parameter estimation but to serve as a prior process for efficiently identifying the time windows of signals and those which may contain new-type signals. Our model also enables the possibility for nearly real-time detection, which is unlikely to be feasible for the usually time-consuming matched-filtering search.

We organize this paper as follows. In Section~\ref{BNN}, we introduce our BNN-based method, including the model architecture, data preparation, training procedure, and the flagging strategy. In Section~\ref{result}, we demonstrate with uncertainty estimation the capability of our model in detecting gravitational waves, first with benchmarking tests and then against the real data from LIGO. In Section~\ref{discussion} we discuss several critical issues in our model and its potential for event prediction. Finally we conclude our work in Section~\ref{conclusion}.

\section{Bayesian Neural Network for Gravitational Wave Detection}
\label{BNN}

\subsection{Bayesian Neural Network}
The Bayesian Neural Network \cite{MacKay:1992:PBF:148147.148165, Hinton1995BayesianLF} is the Deep Neural Network in which its hidden units use probability distributions, instead of point values, as weights and biases. The BNN can provide uncertainty estimation about its prediction and handle overfitting. The parameters in a BNN model are initialized with prior distributions $\textit{p}(W)$, where $W$ represents the model weight. When trained with a given dataset $\mathcal{D}$, these prior distributions are updated to posterior distributions $\textit{p}(W|\mathcal{D})$ via the Bayes' Rule:
\begin{equation}\label{BayesRule}
\textit{p}(W|\mathcal{D}) = \frac{\textit{p}(\mathcal{D}|W)\textit{p}(W)}{\textit{p}(\mathcal{D})},
\end{equation}
where $\textit{p}(\mathcal{D}|W)$ is the likelihood and $\textit{p}(\mathcal{D})$ is the model evidence. One can then obtain the predictive distribution, $\textit{p}(\textbf{y}^*|\textbf{x}^*,\mathcal{D})$, for a new input $\textbf{x}^*$ as
\begin{equation}\label{BayesInference}
\textit{p}(\textbf{y}^*|\textbf{x}^*,\mathcal{D}) = \int\textit{p}(\textbf{y}^*|\textbf{x}^*,W)\textit{p}(W|\mathcal{D})dW.
\end{equation}

However, for most of the modern neural networks the posterior $\textit{p}(W|\mathcal{D})$ cannot be analytically calculated or efficiently sampled due to the enormous number of parameters in the model. To tackle this problem we employ the Variational Inference (VI) \cite{Graves:2011:PVI:2986459.2986721, Jordan1999}, which approximates the true posterior $\textit{p}(W|\mathcal{D})$ with some tractable distributions $\textit{q}_{\bm{\theta}}(W)$ that can be fully parametrized by $\bm{\theta}$. For example, the commonly used Gaussian distribution can be parametrized by the mean $\bm{\mu}$ and standard deviation $\bm{\sigma}$.
The training goal is to minimize the negative evidence lower bound (ELBO): \cite{Graves:2011:PVI:2986459.2986721,2015arXiv150505424B}
\begin{equation}\label{lossFun}
\mathcal{F}(\bm{\theta}) = -\underset{ W\sim\textit{q}_{\bm{\theta}}}{\mathbb{E}}\left[\log\textit{p}(\mathcal{D}|W)\right] + \text{D}_\text{KL}(\textit{q}_{\bm{\theta}}||\textit{p}),
\end{equation}
where the first term is the expectation value of negative log likelihood and the second term is the Kullback-Leibler divergence \cite{kullback1951,kullback1997information} between distributions $\textit{q}_{\bm{\theta}}$ and $\textit{p}$. In classification tasks, the first term is equivalent to the cross-entropy loss.

Within the framework of currently available machine learning tools, the VI estimator used in BNNs is usually achieved by perturbing the weights and biases \cite{Graves:2011:PVI:2986459.2986721,2015arXiv150505424B,2018arXiv180304386W,2013arXiv1312.6114K}. At each forward pass, the model weights and biases are randomly sampled from the distributions $\textit{q}_{\bm{\theta}}(W)$. If the distributions are independent Gaussian distributions, the gradients can be computed using back propagation \cite{2013arXiv1312.6114K}. Thus the predictive distributions can be approximated by propagating the input $\textbf{x}^*$ through the model multiple times and then taking the average:
\begin{align}
\label{eq:mc_integral}
\textit{p}(\textbf{y}^*|\textbf{x}^*,&\mathcal{D}) \approx \int_{W}\textit{p}(\textbf{y}^*|\textbf{x}^*,W)\textit{q}_{\bm{\theta}}(W)dW\nonumber\\
&\approx \frac{1}{N}\sum^{N}_{i=1}\textit{p}(\textbf{y}^*|\textbf{x}^*,W_i),\ W_i \underset{i.i.d.}{\sim} \textit{q}_{\bm{\theta}}(W),
\end{align}
where $N$ is the number of Monte-Carlo samples. 
However, for a mini-batch training the samples in a batch usually share the same weight perturbation for computational efficiency, and this will induce correlation between the gradients and thus make the model hard to converge due to the high variance in the gradient estimation \cite{2018arXiv180304386W}. In addition, it is hard to conduct inference for unknown samples of shared weight perturbations because we are forced to propagate one sample through the model at a time in renewing the weight perturbation. To cope with this, Ref.\cite{2018arXiv180304386W} proposed a flipout estimator that applies a random sign matrix on the weight perturbation matrix for each sample, so as to achieve a pseudo-independent weight sampling for each sample in a mini batch. The flipout estimator does not limit the variance reduction effect during training when trained with a large batch, and it also enables the Monte-Carlo (MC) sampling with mini-batch prediction and thus speeds up the prediction process. In this work we use the Bayes by Backprop (BBB) VI method proposed in Ref.\cite{2015arXiv150505424B} in combination with this flipout estimator \cite{2018arXiv180304386W} to train our model.

Uncertainty estimation is an important feature of the BNN, making it more robust for unknown input than the deterministic neural networks. For a softmax classifier, the predictive uncertainty can be defined as the covariance of the predictive distribution \cite{NIPS2017_7141,Kwon2018UncertaintyQU}:
\begin{equation}
\label{eq:uncertainty_matrix}
\textbf{U} = \frac{1}{N}\sum^{N}_{i=1}(\text{diag}(\textbf{p}_i)-\textbf{p}^{\otimes 2}_i) + \frac{1}{N}\sum^{N}_{i=1}{(\textbf{p}_i - \bar{\textbf{p}})}^{\otimes 2},
\end{equation}
where 
$\textbf{p}_i$ is the predictive vector of the $i$-th MC sample, 
$\bar{\textbf{p}}$ is the mean predictive vector, 
and for a given vector ${\bf v}$ we define the notations
${\bf v}^{\otimes 2} = {\bf v}{\bf v}^{\rm T}$ and
$\text{diag}({\bf v})$ being a diagonal matrix with elements from ${\bf v}$. The first term in Equation \eqref{eq:uncertainty_matrix} is called the aleatoric uncertainty, which captures inherent randomness of the prediction $\textbf{p}_i$. The second term is called the epistemic uncertainty, which originates from the variability of $W$ given the dataset $\mathcal{D}$. In binary classification the off-diagonal elements in $\textbf{U}$ normally provide no useful information so we focus only on the diagonal elements, which are the variances of the predictions in each class. 
Therefore the uncertainty of a prediction on the $k$-th class can be simplified as \cite{Deodato824862}:
\begin{equation}
\label{eq:kthuncertainty}
{u}_k = \frac{1}{N}\sum^{N}_{i=1}(p_{k,i}-p_{k,i}^2) + \frac{1}{N}\sum^{N}_{i=1}(p_{k,i}-\bar{p}_k)^2.
\end{equation}
Because empirically we have ${u}_k < 0.25$, we intuitively define a confidence score for the $k$-th class as \cite{Deodato824862}:
\begin{equation}\label{eq:conf_score}
c_k = 1 - 2\sqrt{{u}_k},
\end{equation}
which indicates how confident the model is in its prediction.

\subsection{Architecture of our Model}

It would be extremely useful if a neural network model for GW detection could also provide information about the duration of a coalescence event, which normally lies between less than a second and tens of seconds depending on the component masses.
With the duration information, we would be able to dramatically reduce the search space for component masses and thus speed up the match filtering process.
Recent work in Ref.~\cite{GebKilParHarSch17} and Ref.~\cite{Gebhard:2019ldz} proposed fully convolutional neural networks based on the structure of WaveNet \cite{2016arXiv160903499V} that accepts input data of various lengths but their model only triggers the alarm around the GW signal peak while providing no information about the signal duration. Here we propose a model that could provide the duration information.

Our approach is a CLDNN model \cite{7178838}, 
which takes advantage of the complementarity of CNNs, LSTMs and DNNs by combining them into one unified architecture (see Sec.~\ref{intro} for full names).
The LSTM \cite{doi:10.1162/neco.1997.9.8.1735,doi:10.1162/089976600300015015} is a type of Recurrent Neural Networks (RNNs) \cite{Rumelhart1986}. An RNN layer consists of several cells, where the hidden units in the current time step have additional recurrent connections to those in next time step \cite{Goodfellow-et-al-2016}. The output of these recurrent connections is called the `hidden state', which enables the cell  to keep the memories from the previous time step. The main function of the LSTM layer is to use an additional `cell state' to further track the long-term memories.
This characteristic of the LSTM enables us to detect not only the peak signal of a coalescence event but also the earlier signals from the inspiral stage. 
We also incorporate the advantage addressed in Ref.~\cite{Gebhard:2019ldz} that when one chooses the half-length of the CNN time window to be the stride size in a sliding-window approach, one could always capture the main part of the waveform in the following time windows even if it lies only partially within the current time window, and in turn the LSTM structure can correlate these windows to make meaningful predictions.

As a first step to deliver the above, we choose the input format of our model to be the strain data sliced into time windows (time steps) of fixed length, each of which is $50\%$ overlapped with its neighbors. 
The details of data generation will be described in Section \ref{data-preparation}. There is essentially no limit on the number of windows so our model is suitable for data of any length. Nevertheless in each operation the size of windows needs to be fixed due to the structural requirement of the LSTM and the fully-connected layers. 

\begin{figure}[!htb]
	\centering
	\includegraphics[width=0.98\linewidth]{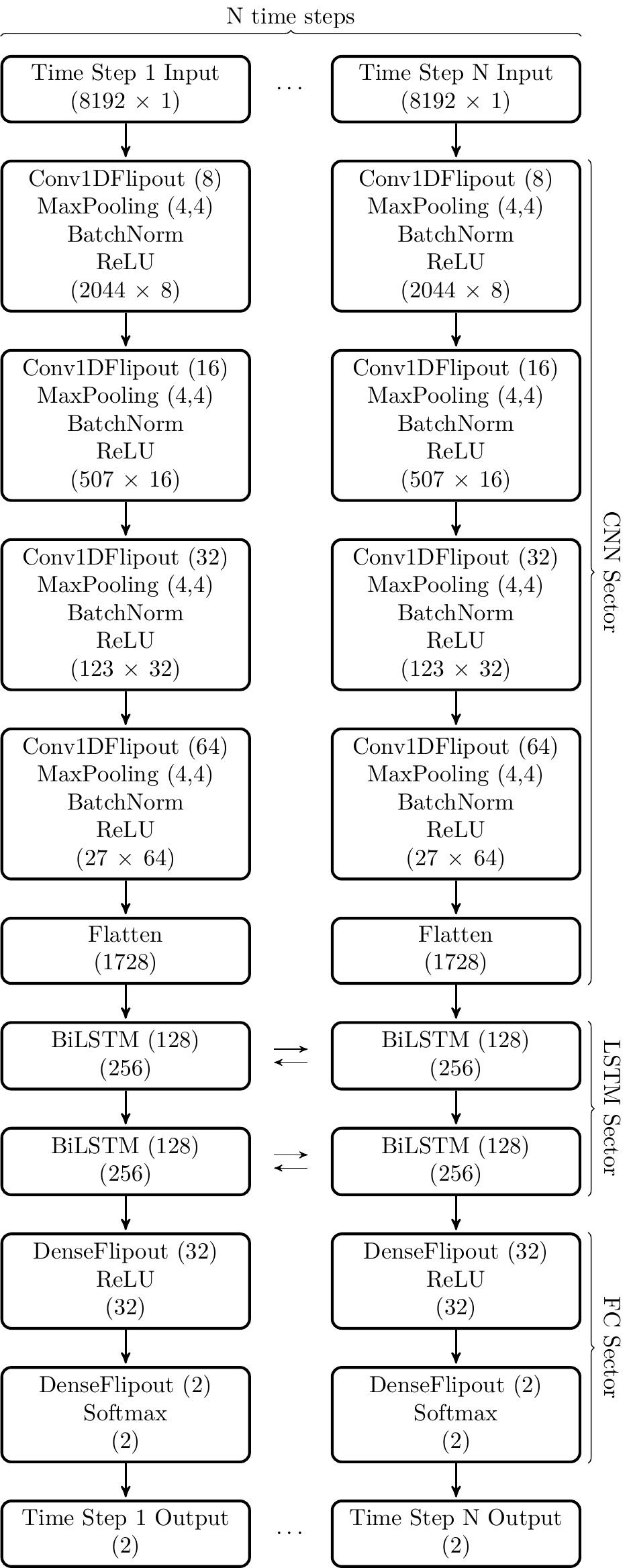}%
	\caption{\label{fig:model}The structure of our Bayesian CLDNN model. The dimensions of the outputs from each block are indicated at the bottom of each block.} 
\end{figure}

Our model is composed of three main sequential sectors: the CNN, LSTM, and Fully-Connected (FC). Fig.~\ref{fig:model} shows the structure.
The CNN sector has four convolutional blocks, each with four layers: Bayesian convolution, 
max pooling \cite{Riesenhuber1999}, 
batch normalization \cite{DBLP:journals/corr/IoffeS15}, and
Rectified Linear Unit (ReLU) activation \cite{Nair2010RectifiedLU}. 
The Bayesian convolution layer uses several filters to extract different features from the input, and each filter has a small set of shared weight distributions called `kernel' to perform the convolution (or cross correlation) \cite{6795724}. Different dilation rates are used to determine the separation of kernel weights in the space. The four Bayesian convolution layers here have 8, 16, 32, and 64 filters, each with the kernel size of 16, 8, 8, and 8 and the dilation rate of 1, 2, 2, and 2, respectively. The stride size of the kernels in all Bayesian convolution layers is fixed to 1.
The max pooling layer helps us down-sample the data by picking up the maximum value in the pooling window. All the pooling layers have a window size of 4, sliding with a stride size of 4. 
After the convolution the data are flattened and then passed into the LSTM sector. 

The LSTM sector has two bidirectional Bayesian LSTM layers, each with 128 hidden units (weights) in each direction. Finally, the FC sector contains two Bayesian FC layers, with 32 and 2 hidden units respectively. The first Bayesian FC layer is followed by an ReLU activation layer, and the second by a softmax activation layer in order to confine the output values between 0 and 1. 
All Bayesian layers employ both the VI \cite{Graves:2011:PVI:2986459.2986721,2015arXiv150505424B,2017arXiv170402798F} and the flipout technique \cite{2018arXiv180304386W} to generate pseudo-independent weight and bias perturbations in the hidden layers.

Our model has a total of around 4.65 million parameters.
Its outputs are the class results (class 0 for noise and class 1 for signal), each attached with a confidence score between 0 and 1 as a reference to judge on the existence of the GW signals from coalescence events in the corresponding time windows. 

\subsection{Data Preparation}\label{data-preparation}

\subsubsection{Real Noise Data}\label{Sec:RND}
In this work we use the LIGO Livingston O2 data segments provided by the Gravitational Wave Open Science Center \cite{Vallisneri:2014vxa} (GWOSC) as our background noise of GW injection. We randomly select 15 data segment files from the first few days of the observation for training\footnote{December 1$^{\rm st}$ to 3$^{\rm rd}$, 2016.}, 5 files from the final day for validation\footnote{August 25$^{\rm th}$, 2017.}, and 8 files from the middle for testing the model\footnote{April 9$^{\rm th}$ and 10$^{\rm th}$, 2017.}. All data segment files are 4096 seconds in length and have a quality of at least \texttt{CBC\_CAT3} = 100. They do not contain any of the events or marginal triggers published by LIGO \cite{LIGOScientific:2018mvr}. We down-sample the sample rate of the strain data from 16384 to 8192 Hz in order to save memory.

\subsubsection{\label{Sec:STGWS}Simulated Templates of GW signals}
We use the software packages \texttt{PyCBC} \cite{pycbc,Usman_2016} and \texttt{LALSuite} \cite{lalsuite} to generate GW signals of quasi-circular, non-spinning BBH coalescence with the effective-one-body model \texttt{SEOBNRv4T} \cite{Bohe:2016gbl}. The signals are simulated with a sample rate of 8192 Hz and a cut-off frequency of 20 Hz. The choice of component masses is similar to those in Ref.~\cite{George:2016hay, George:2017pmj}. 
For the training dataset the BBH component masses range between $5 M_\odot$ and $75 M_\odot$ in steps of $1 M_\odot$, with a mass ratio of $q = M_1/M_2 \leq 10$. 
For the validation dataset the masses are offset by 0.5 $M_\odot$ with respect to those in the training dataset.
This offset helps ensure that the network is not overfitting by memorizing only the inputs shown to it without learning to generalize to new inputs \cite{George:2016hay, George:2017pmj}.
For the testing dataset the masses are the collection of those values used in the above two datasets.
In order to make our simulated samples more realistic, we also incorporate the GPS time stamps (of the associated real noise; see below) and a set of randomly generated right ascensions and declinations (for the signal part) into the samples following the convention of LIGO L1 detector. This in turn means that every simulated waveform in our dataset is different.

\subsubsection{Data Generation}
\label{data-generation}
In generating various datasets for training, validation, and testing,
we use the real noise randomly picked up from the LIGO dataset (Sec.~\ref{Sec:RND}) for every sample, and inject the simulated signals (Sec.~\ref{Sec:STGWS}) into half of the samples for each dataset.
To manipulate the signal-to-noise ratio (SNR) for the samples that contain signals, we pre-calculate the power spectral density (PSD) of the noise $S_n(f)$ and tune the strength of the GW signal ${h}$ so that the optimal SNR, $\rho_\text{opt}$, defined by
\cite{Sathyaprakash:2009xs}
\begin{equation}
\rho^{2}_\text{opt} = 4\int\limits^\infty_0df\frac{\left\lvert\tilde{h}(f)\right\rvert^2}{S_n(f)},
\end{equation}%
lies within the desired range (see below).
Here $\tilde{h}(f)$ is the Fourier transform of $h$. 
After the injection we then whiten the data using the PSD re-estimated including the injected signal.
The length of the data in this process is 16 seconds and we keep only the central 8 seconds to avoid FFT artifacts.
We also purposely arrange for the signals to peak within the last 2 seconds of this final 8 seconds, in a hope that our model could detect the coalescence event as soon as it comes into the analyzed data. The 2-second diversity in the signal position is expected to lead to a better sensitivity of our model in event position.
Our tests did show that such arrangement for signals (to peak within the last 2 seconds) for training helped largely not only for a higher sensitivity of early detections but also for a more complete mapping of the waveform.

\begin{figure}[!tb]
	\centering
	\includegraphics[width=1\linewidth]{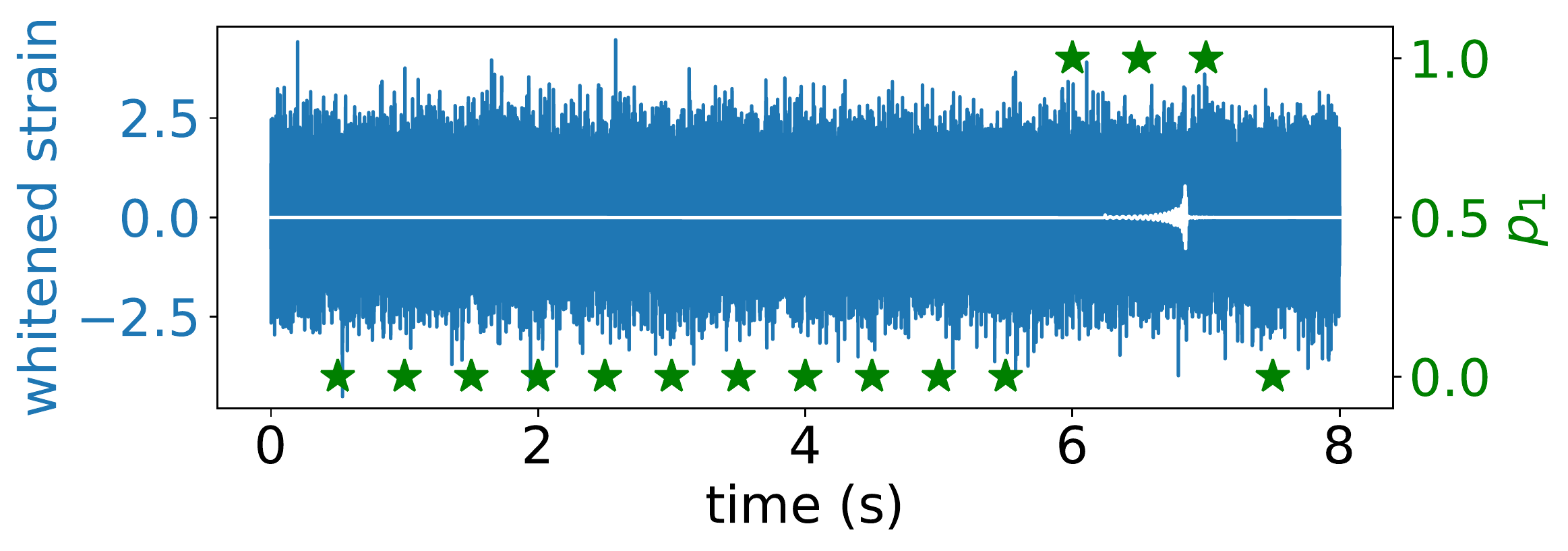}%
	\caption{\label{fig:sample}Simulated LIGO L1 strain data (blue curve) that contain a GW signal (white curve) with $m_1=45.0$, $m_2=38.5$, $\rho_\text{opt}=8$, right ascension $\pi/4$ and declination $\pi/4$. The green stars are the marks (0 or 1) of the time windows centered at each window.}
\end{figure}

We then slice and standardize the whitened data into 15 time windows of 1 second, each with 0.5-second overlaps with its neighbors. Finally, we mark a time window with 1 if it contains a GW signal of longer than 0.25 seconds, or longer than half of the  signal duration when the duration is shorter than 0.5 second. All other windows are marked with 0. An example of the simulated data sample is shown in Fig.\ref{fig:sample}.

To carry out this work, we generate 40960 and 4096 samples for the training dataset and the validation dataset respectively, each with half of the samples containing signals of SNRs randomly drawn between 5 and 15 in steps of 0.5.
Similarly, for the testing dataset (used in Sec.~\ref{Sec:perf-ag-mock-data} to quantify the performance of our model), we generate 512 samples, half of which contain signals to cover an SNR range from 2 to 18 in steps of 0.5. For this testing performed in Sec.~\ref{Sec:perf-ag-mock-data}, we did not go for a larger dataset beyond the size of 512 because it took us about five days while already giving good quantitative results.

\subsection{Training Procedure}
To deliver our model, we employ the tools provided by the  software packages TensorFlow \cite{tensorflow2015-whitepaper} and the TensorFlow Probability \cite{2017arXiv171110604D}. For the Bayesian LSTM layers we build a customized Bayesian LSTM cell that applies the weight perturbation and the flipout estimator on the hidden kernel, recurrent kernel and bias. The LSTM cell is then wrapped by the ordinary RNN layer functionality provided in TensorFlow. For updating the model weights during training, we use the ADAM \cite{2014arXiv1412.6980K} optimizer with a scheduled learning rate of
\begin{equation}\label{eq:lr}
\text{lr}(e) = \begin{cases}
	0.005, & \text{if } e \leq 15;\\
	0.005*\exp(0.05*(15-e)), & \text{otherwise};
	\end{cases}
\end{equation}%
where $e$ is the number of training epoch. We use a combination of the sparse categorical cross entropy and the KL divergence as our loss function. Because in our training scheme the KL divergences of each layer and cell are accumulated into the regularization loss whenever passing through the data, we have to weigh the KL divergence in each layer and cell with the value of $\frac{1}{CN}$ \cite{2017arXiv170402798F}, where $C$ is the number of time steps in a sample and $N$ is the total number of samples for training, so that it is applied only once per epoch. Otherwise the accumulated large KL divergence will over-regularize the model and eventually stop the model from learning.

About the hardware, we train our model of 150 epochs and a batch size of 256 on a computer with Intel\textsuperscript{\textregistered} Core\textsuperscript{\texttrademark} i5 6500 CPU and AMD\textsuperscript{\texttrademark} RX 480 GPU.
Even with such moderate computational power
and a code of Bayesian LSTM cell unoptimized for GPU,
the whole training process normally takes only about 50 hours.
This is a critical feature of our work. 
After the last epoch we save only the model parameters because we want to minimize the KL loss, which keeps decreasing during the  training.
 
\subsection{\label{sec:flags}Flagging Strategy}

Our model performs tasks of binary classification,
with a predictive value of $p_1$ indicating the identification for GW signals and a value of $p_0$ for noise, where $p_0+p_1=1$.
Thus we need to keep only $p_1$ for subsequent analysis. 
We also calculate the average predictive value $\bar{p}_1$ and the confidence score $c_1$ for $\bar{p}_1$ using Equations \eqref{eq:mc_integral} and \eqref{eq:conf_score} respectively.

Considering the issue of confidence level in statistics, we weigh the $\bar{p}_1$ with its confidence score $c_1$.
This is to avoid our model from accepting or rejecting a time window for signals when the confidence score $c_1$ is low.
Thus we flag all the time windows with three distinct states: \textit{trigger}, \textit{noise}  and \textit{awareness}.
A time window $t$ is flagged with a trigger state when the trigger score defined as
\begin{equation}
\label{eq:trigger_score}
s_t = \bar{p}_{1,t}\times c_{1,t},
\end{equation}
is larger than 0.5,
where $\bar{p}_{1,t}$ and $c_{1,t}$ are the $\bar{p}_1$ and $c_1$ of the time window $t$ respectively.
Such $s_t$ can be regarded as the significance level of a detection.

Similarly a time window $t$ is flagged with a noise state when the noise score defined as
\begin{align}
\label{eq:noise_score}
n_t &= \bar{p}_{0,t} \times c_{0,t}\;,\nonumber\\
&= (1-\bar{p}_{1,t})\times c_{1,t}\;,
\end{align}
is larger than 0.5. 
We note that in a system of binary classification, $c_{0,t}$ equals $c_{1,t}$.
We also note that $s_t+n_t=c_{1,t}$.

Those time windows that are not flagged with a trigger state nor a noise state will be flagged with an awareness state. 
The role of such a state can be illustrated in a Bayesian dog-null classifier for images when we feed it with an image of cat.
It is likely that the classifier will give a reasonable $p_1$ but low $c_1$. In such a case we can only say that the model `notices' something though not sure what it is.

\begin{figure}[!tb]
	\centering
	\includegraphics[width=1\linewidth]{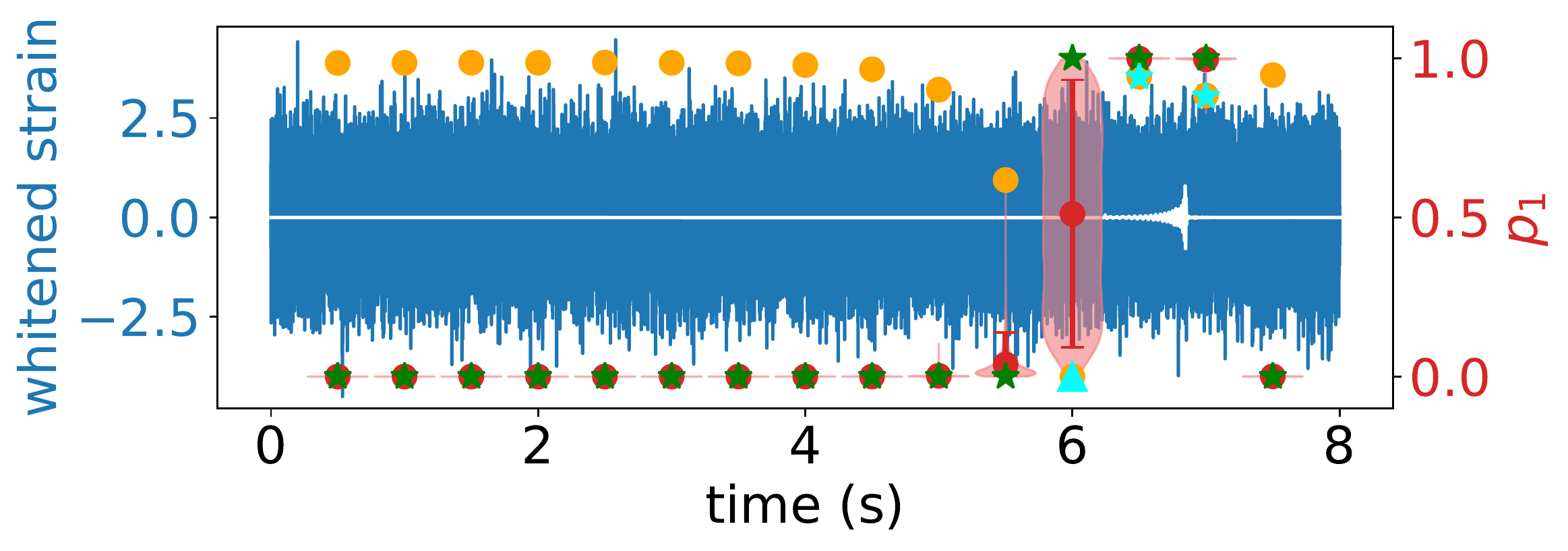}%
	\caption{\label{fig:sample_pred1}The prediction with flagging results from our model for the input data presented in Fig.~\ref{fig:sample}.
	The three cyan markers indicate the trigger scores $s_t$ (Eq.~\eqref{eq:trigger_score}), where two trigger states are shown as stars and one awareness state as a triangle.
	We also show the confidence score $c_{1,t}$ (organge dots), averaged probability for being a signal $\bar{p}_{1,t}$ (red dots), distribution of $p_{1,t}$ (pink areas), and the 90\% intervals of $p_{1,t}$ (red error bars), all based on a Monte-Carlo process with a sampling size of 4096.}
\end{figure}

As a demonstration we apply the above flagging strategy to the data presented in Fig.~\ref{fig:sample} and the results are shown in Fig.~\ref{fig:sample_pred1}. 
Because the flipout estimator enables batch prediction, we use a batch size of 32 to dramatically accelerate the MC sampling process (see Sec.~\ref{size-of-MC-sampling}). 
As shown in Fig.~\ref{fig:sample_pred1}, among the three time windows that are marked with 1 (green stars) for indicating the GW singles, our model flag two with trigger states (cyan stars) and one with an awareness state (cyan triangle) likely due to the much weaker signal within this time window.
This demonstrates the usefulness of the awareness state in practice, which has actually incorporated the sliding-window search.
In Fig.~\ref{fig:sample_pred1} we also show the confidence score $c_{1,t}$ (organge dots), the averaged predictive probability for being a signal $\bar{p}_{1,t}$ (red dots), the distribution of $p_{1,t}$ (pink areas), and the 90\% intervals of $p_{1,t}$ (red error bars), all obtained from a set of data sampled for 4096 times.

It is obvious that the awareness state is associated with a low confidence score $c_{1,t}$, which indicates a large uncertainty in a prediction.
It suggests the need for further investigations on this particular time window.
In practice, we could ignore these awareness states if we have had trigger states next to them because the purpose of our model is to detect signal events.
However if the awareness states do not come with any neighboring trigger states, we could pursue further investigations in order not to miss any detection opportunities for signals of even unknown types.
In a training process, on the other hand, these awareness states together with their known markers can be further incorporated into a retraining process, so that the retrained model could become more capable in discriminating between the signals and the noise rather than putting either into the awareness state.

\section{Results}\label{result}

\subsection{Quantitative Tests for Performance}
\label{Sec:perf-ag-mock-data}

To quantify the performance of our model, we take four different measures, namely the TPR, TER, FPR, and FER as defined below, against the dataset generated in Section \ref{data-preparation}, which contains real noise and simulated signals.
The MC sample size in obtaining $\bar{p}_1$ and $c_1$ is 4096. 
Because our model classifier gives three states instead of two,
we cannot apply the commonly used confusion matrix to obtain the sensitivity or the false positive rate for our results.

\begin{figure}[!htb]
	\centering
	\includegraphics[width=1\linewidth]{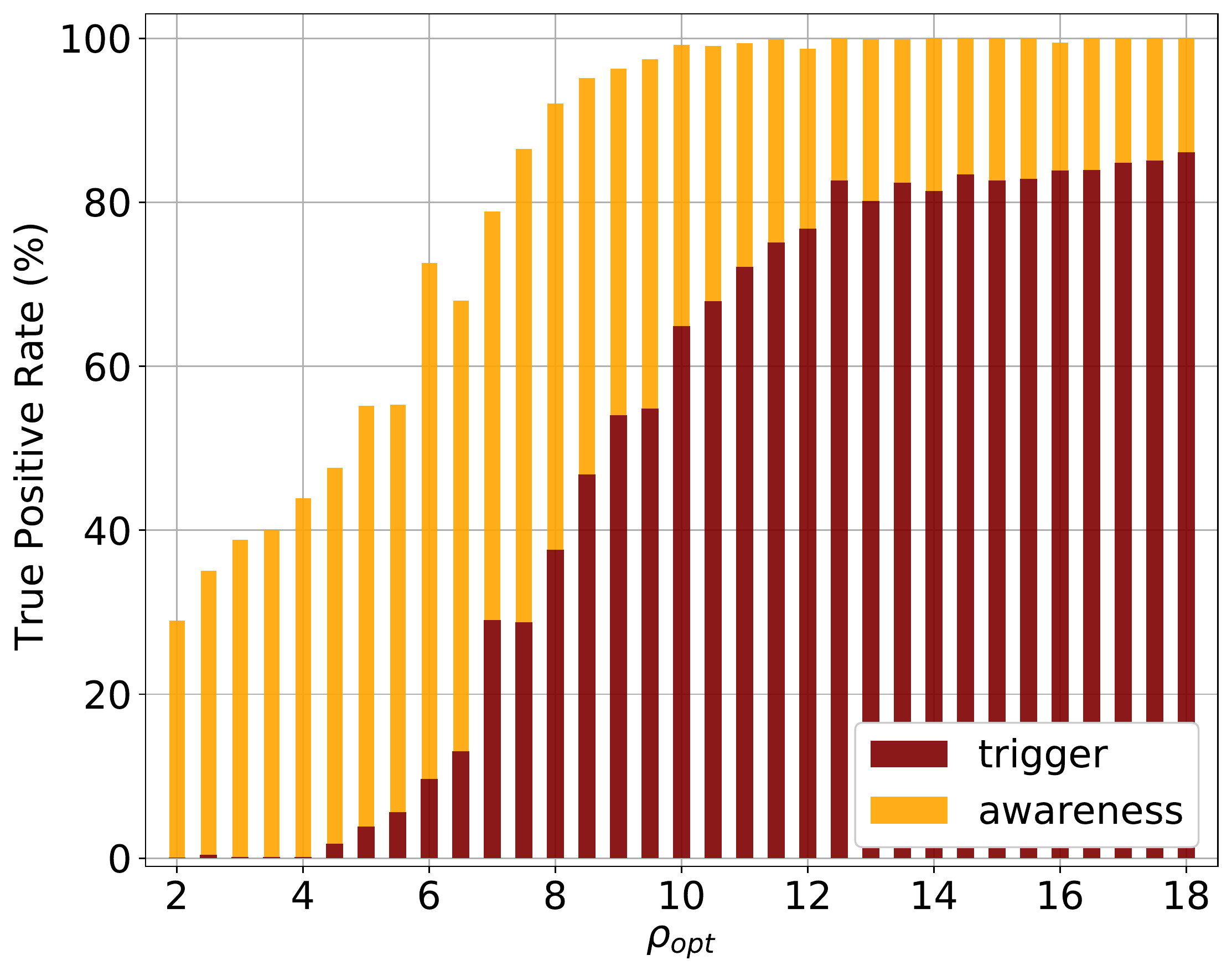}%
	\caption{\label{fig:wave_sens}The True Positive Rate (TPR) as a function of the SNR $\rho_\text{opt}$ in the performance test of our model. The TPR reaches 90\% when $\rho_\text{opt}>8$.} 
\end{figure}

The first measure is the `true positive rate' (TPR), which is the ratio of the trigger states plus the awareness states among all the time windows marked with `1' (with GW signals). To be precise, among all the time windows marked with 1, if the numbers of states for trigger, awareness, and noise are $N_{\rm trg}$, $N_{\rm aw}$, and $N_{\rm noi}$ respectively, then the TPR is defined as
\begin{equation}
\label{eq:TPR}
\text{TPR} = \frac{N_{\rm trg}+N_{\rm aw}}{N_{\rm trg}+N_{\rm aw}+N_{\rm noi}}.
\end{equation}
Such TPR can be regarded as the `waveform sensitivity',
which indicates how well the model can capture the waveform structure. Fig.~\ref{fig:wave_sens} shows the results.
It is clear that for GW signals with an SNR of $\rho_\text{opt}>8$, our model detects or is aware of more than 90\% of the time windows in the GW waveforms. The persistent awareness rate of about 20\% at high $\rho_\text{opt}$ is due to the obscuration from the noise at the beginning parts of the waveforms, which are always weaker then the noise.
In principle this rate of 20\% could be reduced if we retrain the model with these marked awareness states.

\begin{figure}[!htb]
	\centering
	\includegraphics[width=1\linewidth]{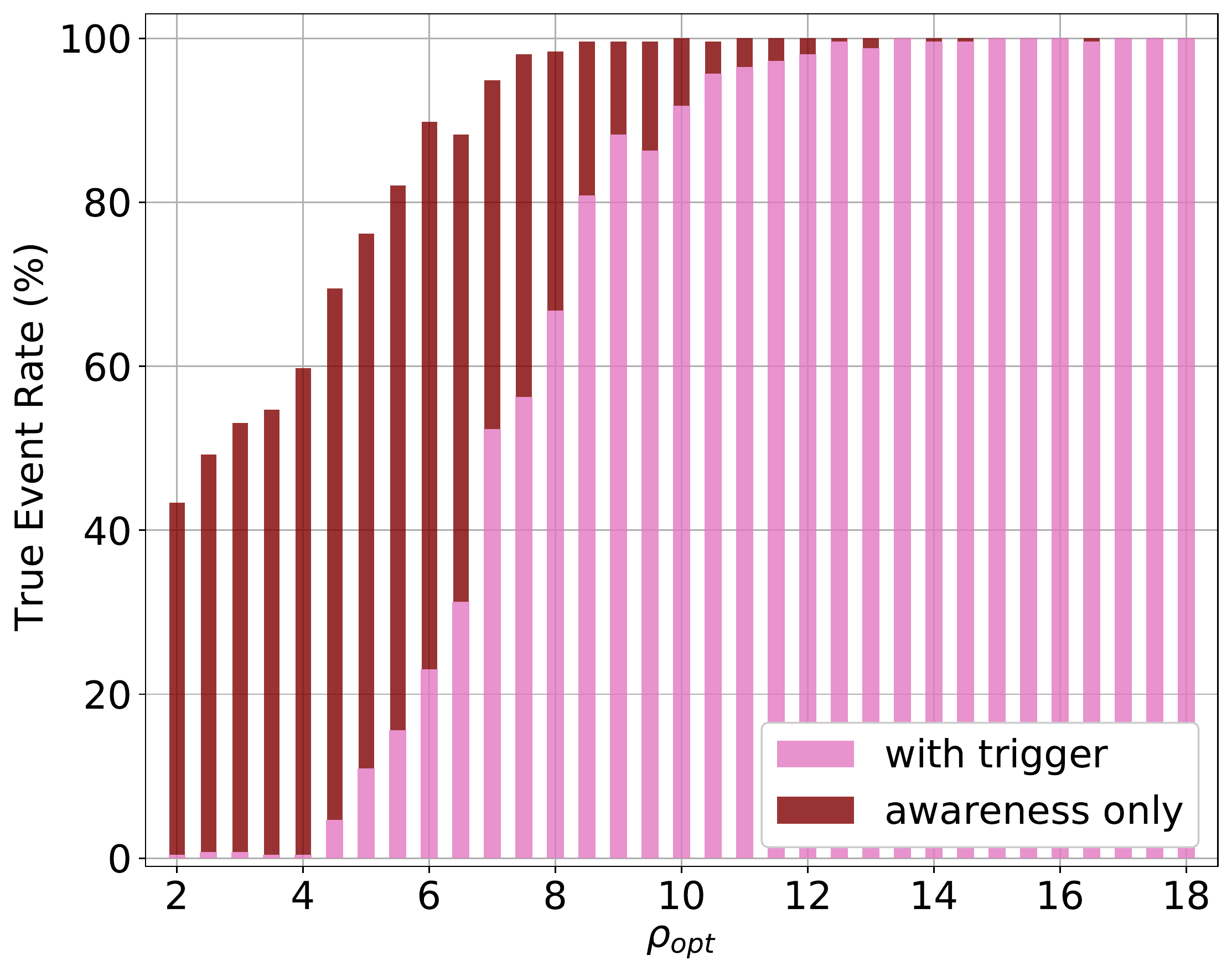}%
	\caption{\label{fig:event_sens}The True Event Rate (TER) as a function of the SNR $\rho_\text{opt}$ in the performance test of our model. The TER reaches 90\% when $\rho_\text{opt}>7$.} 
\end{figure}

The second measure is the `true event rate' (TER), which is the ratio of the events with at least one trigger or awareness among all the events with GW signals.
Among all the signal events, if the numbers of events with `at least one trigger', with `awareness only', and with `noise only' are $E_{\rm trg}$, $E_{\rm aw}$, and $E_{\rm noi}$ respectively, then the TER is defined as
\begin{equation}
\label{eq:TER}
\text{TER} = \frac{E_{\rm trg}+E_{\rm aw}}{E_{\rm trg}+E_{\rm aw}+E_{\rm noi}}.
\end{equation}
Such TER can be regarded as the `event sensitivity', which indicates how well our model can pick up the true events.
Fig.~\ref{fig:event_sens} shows the results.
Our model achieves a TER of 90\% when $\rho_\text{opt} > 7$ and 100\% when $\rho_\text{opt} > 8.5$.
For $\rho_\text{opt} > 11$ we still see few cases with awareness only and our investigation shows that this is due to some outliers in the noise leading to low confidence.

In evaluating the TER and TPR results, it is quite encouraging to see the capability of our model in not only detecting the GW events (TER) but also identifying the full lengths of waveform durations in the events (TPR), although the latter demonstrates a slightly lower sensitivity.
We also note that our model trained for the non-spinning BBHs successfully detected the spinning BBHs in similar statistics.

On the other hand, we could define two similar measures for false detections.
Thus the third measure is the `false positive rate' (FPR), which is the ratio of the trigger states plus the awareness states among all the time windows marked with `0' (without GW signals). In other words, among all the time windows marked with 0, if the numbers of states for trigger, awareness, and noise are $N^0_{\rm trg}$, $N^0_{\rm aw}$, and $N^0_{\rm noi}$ respectively, then the FPR is defined as
\begin{equation}
\label{eq:FPR}
\text{FPR} = \frac{N^0_{\rm trg}+N^0_{\rm aw}}{N^0_{\rm trg}+N^0_{\rm aw}+N^0_{\rm noi}},
\end{equation}
which is the sum of the `false trigger rate' ${N^0_{\rm trg}}/({N^0_{\rm trg}+N^0_{\rm aw}+N^0_{\rm noi}})$ and the `false awareness rate' ${N^0_{\rm aw}}/({N^0_{\rm trg}+N^0_{\rm aw}+N^0_{\rm noi}})$.
This FPR indicates how likely our model will falsely capture the waveform signals. Because the false detections are measured in a background of pure noise, the FPR is not a function of the SNR.
Our model shows a false trigger rate of $0.067\%$ and a false awareness rate of $19.6\%$, summed up to an FPR of about $19.7\%$.
The false trigger rate is unnoticeably small and they are mainly due to the glitches in the noise. 
On the other hand, although the current false awareness rate of $19.6\%$ is unignorable, we find that when conducting the same test using noise datasets of different sizes for training, the false awareness rate decreases monotonically with the size of noise dataset for training. Therefore the current $19.6\%$ is expected be further reduced by a larger noise training dataset, which would contain more variability in the background noise. 
In addition, we should be able to further reject false triggers or awareness when performing the coincidence test (see Sec.~\ref{multi-network-detection}).

The last measure is the `false event rate' (FER), which is the ratio of the events with at least one trigger or awareness among all the null events without GW signals.
Among all the null events, if the numbers of events with `at least one trigger', with `awareness only', and with `noise only' are $E^0_{\rm trg}$, $E^0_{\rm aw}$, and $E^0_{\rm noi}$ respectively, then the FER is defined as
\begin{equation}
\label{eq:FER}
\text{FER} = \frac{E^0_{\rm trg}+E^0_{\rm aw}}{E^0_{\rm trg}+E^0_{\rm aw}+E^0_{\rm noi}},
\end{equation}
which is the sum of the `false event trigger rate' ${E^0_{\rm trg}}/({E^0_{\rm trg}+E^0_{\rm aw}+E^0_{\rm noi}})$ and the `false event awareness rate' ${E^0_{\rm aw}}/({E^0_{\rm trg}+E^0_{\rm aw}+E^0_{\rm noi}})$.
Our model shows a false event trigger rate of $0.485\%$ and a false event awareness rate of $47.2\%$, summed up to an FER of about $47.7\%$.
This shows that while our model is capable of detecting the true events as previously seen, it falsely over-predicts the signal events. Again this is simply due to the smallness of our noise dataset for training and can be improved by enlarging the noise training dataset as well as going deeper in the training.

\subsection{Performance against Real Events}
\label{performance-against-real-data}

In this section we test our model with the LIGO Livingston O2 data that contain confidence detections of BBH coalescence events \cite{PhysRevLett.118.221101,Abbott_2017,LIGOScientific:2018mvr,Abbott_2019,PhysRevLett.119.141101}. 

We downloaded the 32-second strain data from GWOSC, down-sampled them to 8192 Hz, calculated the PSD, whitened the data, and then took the 8-second chunks that contain the events in their last seconds. As in the training process, each chunk of the real data here was also sliced into 15 one-second windows with a stride of 0.5 second. For each chunk of the event data, we performed an MC sampling of size 4096, with a batch size of 32. 
When operated on a moderate GPU-equipped PC, the whole process of flagging took only about 20 seconds for each chunk.
Thus our model should be able to achieve nearly real-time detection if we employ better hardware and optimize the Bayesian LSTM layer in gaining the full GPU support. Shorter data chunks and a smaller size of the MC sampling should also help on this.

\begin{figure*}[!htb]
	\thisfloatpagestyle{empty}
	\centering
	\subfloat{\includegraphics[clip,width=\linewidth]{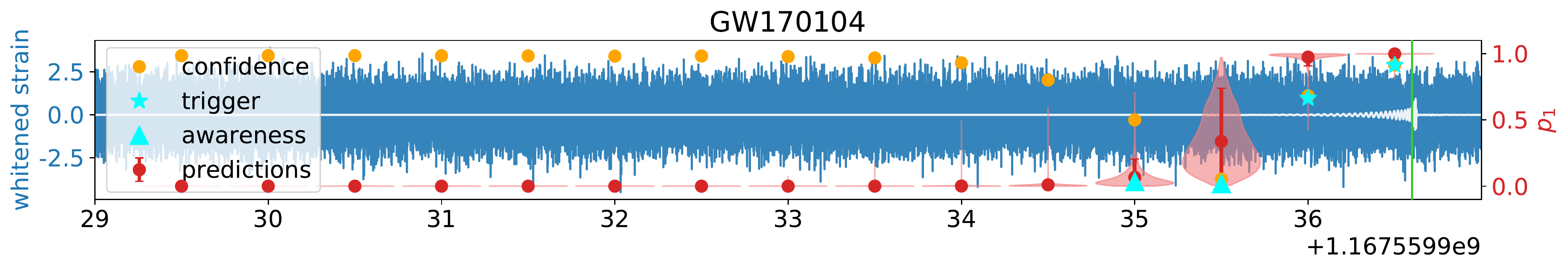}}\\[-5.3pt]
	\subfloat{\includegraphics[clip,width=\linewidth]{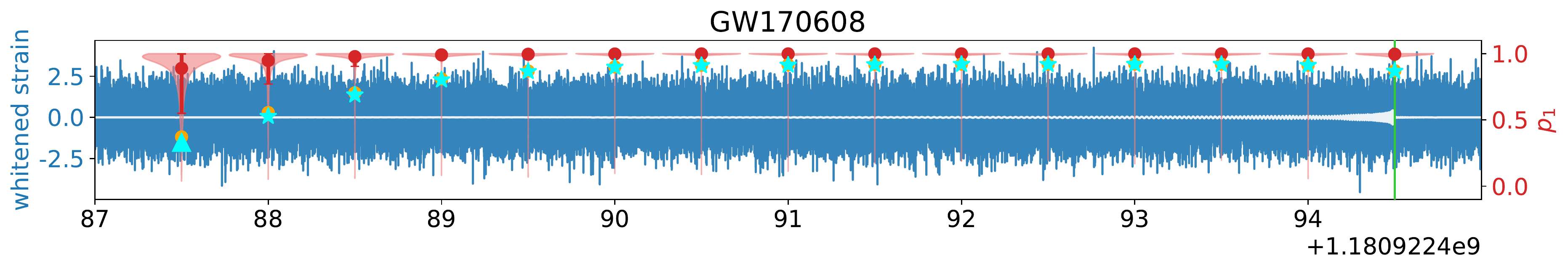}}\\[-5.3pt]
	\subfloat{\includegraphics[clip,width=\linewidth]{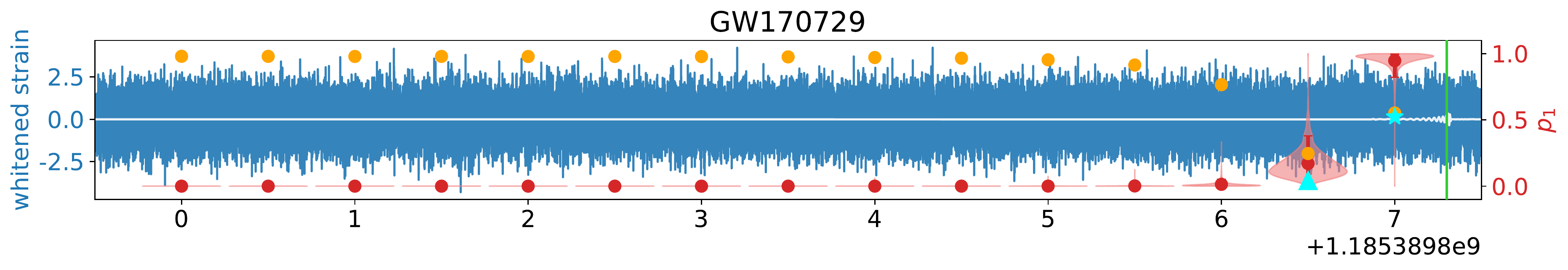}}\\[-5.3pt]
	\subfloat{\includegraphics[clip,width=\linewidth]{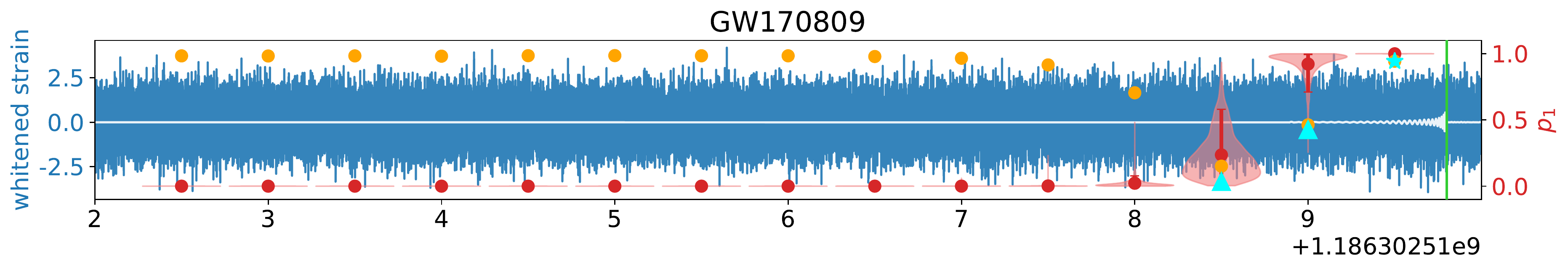}}\\[-5.3pt]
	\subfloat{\includegraphics[clip,width=\linewidth]{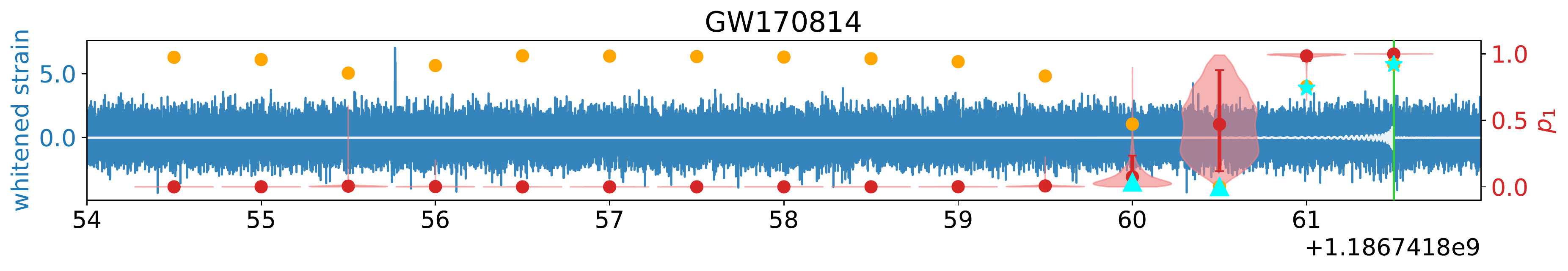}}\\[-5.3pt]
	\subfloat{\includegraphics[clip,width=\linewidth]{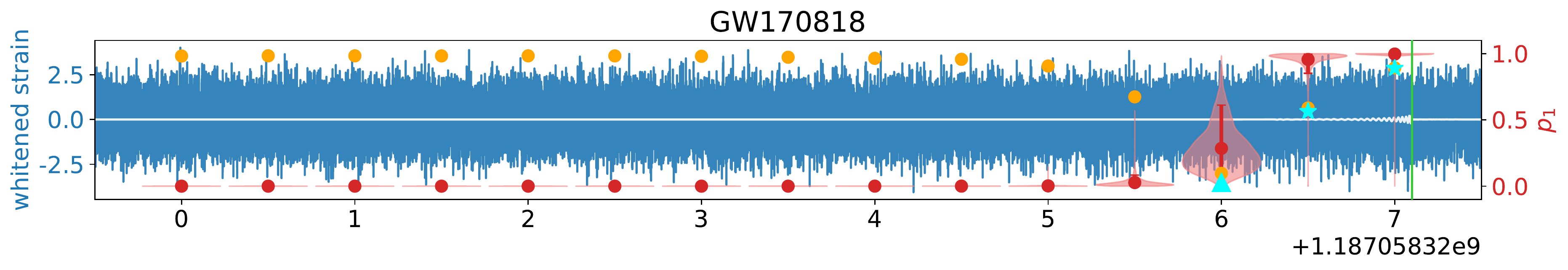}}\\[-5.3pt]
	\subfloat{\includegraphics[clip,width=\linewidth]{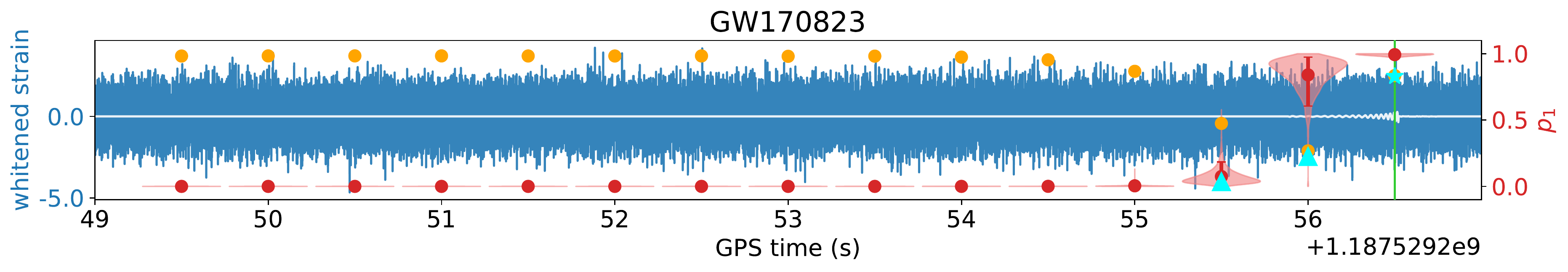}}\\[-5.2pt]
	\caption{\label{fig:LIGO_BBH}Successful flagging results of our model for the LIGO Livingston O2 data that contain GW signals from BBH coalescence events.
		The green vertical lines indicate the event times.
		Other symbols and colors in these plots follow the same definitions as in Fig.~\ref{fig:sample} and Fig.~\ref{fig:sample_pred1}.}
\end{figure*}

The results for the BBH coalescence events are shown in Fig.~\ref{fig:LIGO_BBH}. The symbols and colors in these plots follow the same definitions as in Fig.~\ref{fig:sample} and Fig.~\ref{fig:sample_pred1}. For reference purpose, we also plot the whitened GW waveforms (white curves) reconstructed from the values of component masses, luminosity distances, right ascensions and declinations provided in Ref.~\cite{LIGOScientific:2018mvr} using the \texttt{SEOBNRv4T} model. 

It is clear that our model successfully detected all the events. 
It is also important to note that our model successfully triggered all the time windows that contained the GW170608 signal, which spanned a period of nearly 7 seconds. This demonstrates the capability of our model in capturing the full length of a long-duration GW signal.
While in literature the matched-filtering search detected GW170729 with rather high false alarm rate \cite{Abbott_2019,LIGOScientific:2018mvr},
our detection of GW170729 also comes with a relatively lower confidence score.
This is likely due to the noise fluctuation in the background and could be improved by deepening the model or increasing the size of the training dataset.

\begin{figure*}[!htbp]
	\centering
	\includegraphics[width=1\linewidth]{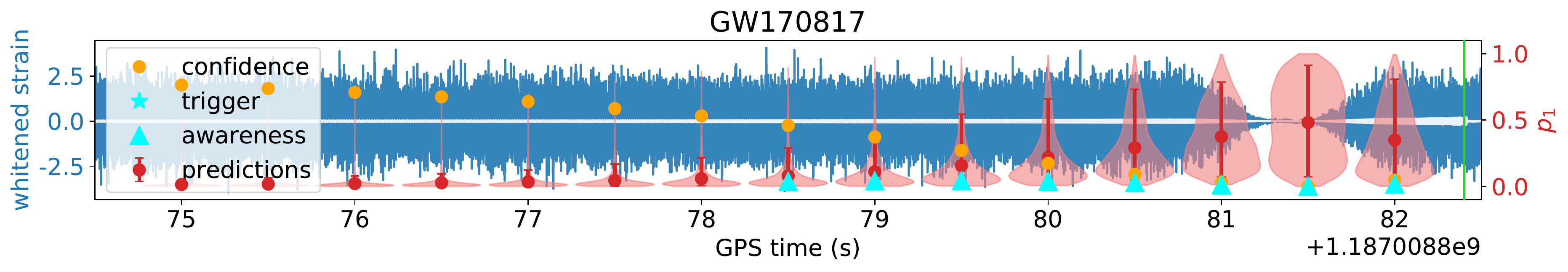}%
	\caption{\label{fig:LIGO_BNS}The flagging results of the LIGO Livingston O2 data set GW170817, which contains a BNS coalescence event. Here we have applied an inverse Tukey window \cite{TheLIGOScientific:2017qsa,Pankow:2018qpo} to mitigate the glitch that occurred about 1.1 second before the event. The white curve is the reconstructed signal of the related BNS waveform.} 
\end{figure*}%

Although our model was trained with only the BBH coalescence events, we experimentally tested our model against a binary-neutron-star (BNS) coalescence event, the GW170817 \cite{TheLIGOScientific:2017qsa}, which has masses of 1.27~$ M_\odot$ and 1.46~$M_\odot$. Fig.~\ref{fig:LIGO_BNS} shows the result. For this particular set of Livingston O2 data where there was a large glitch about 1.1 seconds before the event \cite{TheLIGOScientific:2017qsa}, we employed a method used for the rapid reanalysis in Ref.~\cite{TheLIGOScientific:2017qsa,Pankow:2018qpo} to mitigate this large glitch. 
The method applies an inverse Tukey window to zero out the data around the glitch.
For reference purpose we also plotted the reconstructed BNS GW waveform (white curve) using the \texttt{TaylorT4} model \cite{Boyle:2007ft}.
As shown in Fig.~\ref{fig:LIGO_BNS}, our model was aware (with no triggers) of the time windows that contained the waveform of the  BNS event.

\begin{figure}[!tb]
	\centering
	\includegraphics[width=\linewidth]{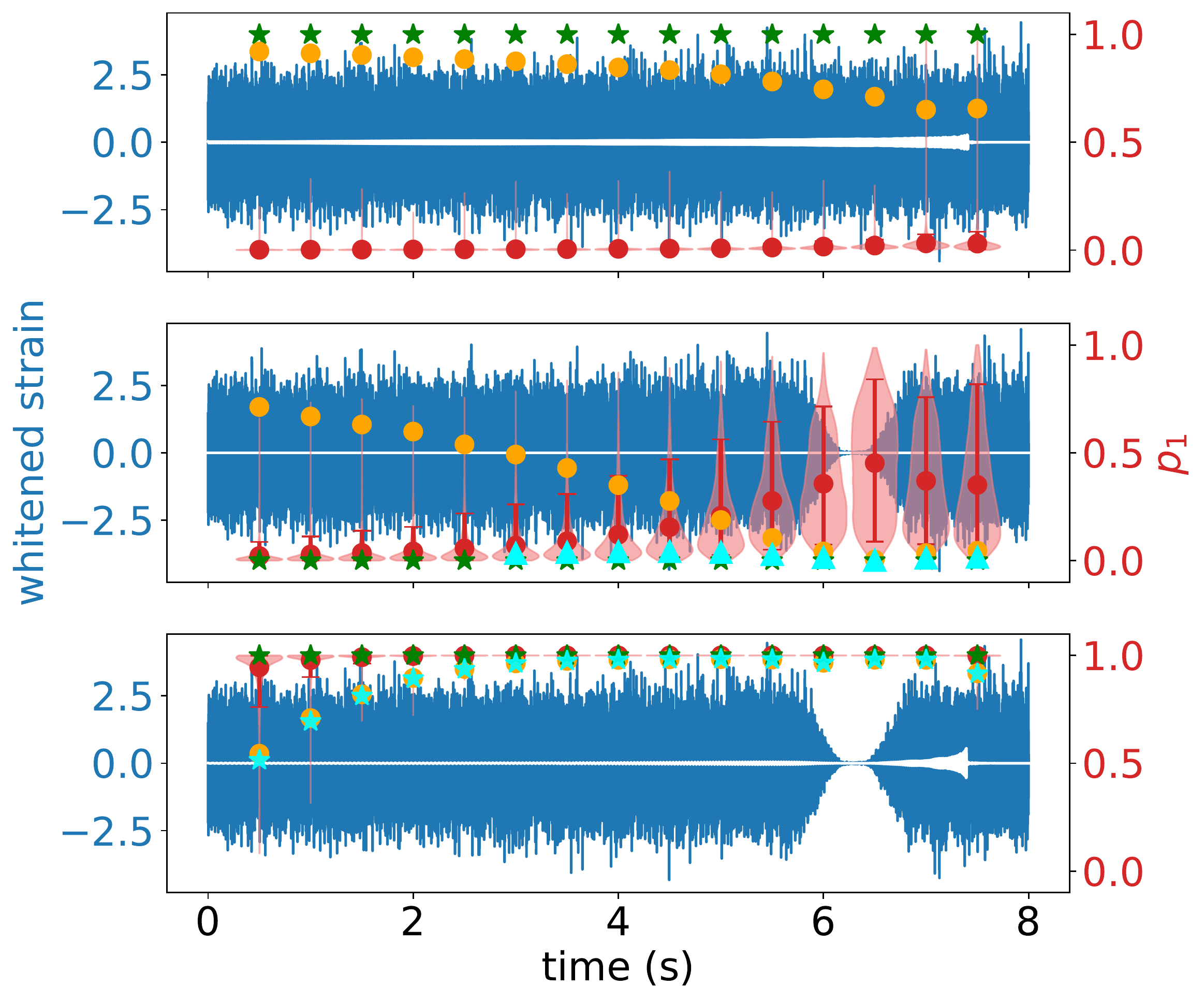}%
	\caption{\label{fig:170817val}Three runs for cross validation to test the influence of a Tukey window on the GW170817 result in Fig.~\ref{fig:LIGO_BNS}.
	With the same normal noise background from the LIGO data, the first run contained a BNS waveform same as shown in Fig.~\ref{fig:LIGO_BNS} to mimic the GW170817, without any window treatment (top panel); the second run contained no waveform but applied with an inverse Tukey window (middle panel); the third contained a BBH waveform to mimic a loud BBH event, with an inverse Tukey window (bottom panel).} 
\end{figure}

To quantify how much our result had been affected by the Tukey window, which actually deformed the strain data and can be thought as an adversarial example \cite{2013arXiv1312.6199S,8294186}, we performed the following three runs for cross validation (see Fig.~\ref{fig:170817val}).
With the same normal noise background (without glitches) from the LIGO Livingston O2 data, the first run contained a BNS waveform same as shown in Fig.~\ref{fig:LIGO_BNS} to mimic the GW170817, without any window treatment (top panel in Fig.~\ref{fig:170817val}); the second run contained no waveform but applied with an inverse Tukey window (middle panel in Fig.~\ref{fig:170817val}); the third contained a BBH waveform to mimic a loud BBH event ($m_1=8.5\ M_\odot$, $m_2=6.0\ M_\odot$, and $\rho_\text{opt}=12$), with an inverse Tukey window (bottom panel in Fig.~\ref{fig:170817val}).
It is evident that our model rejected the GW170817-like signal in the top panel, raised awareness for the Tukey window in the middle panel, and raised triggers for the BBH event in the bottom panel without being affected by the Tukey window.
We also note that the $\bar{p}_1$ in the middle panel exhibits a behavior similar to the one shown in Fig.~\ref{fig:LIGO_BNS}.
Therefore it is likely that the awareness seen in the GW170817 (Fig.~\ref{fig:LIGO_BNS}) is due to the treatment involving the Tukey window.

\begin{figure}[!hbt]
	\centering
	\includegraphics[width=\linewidth]{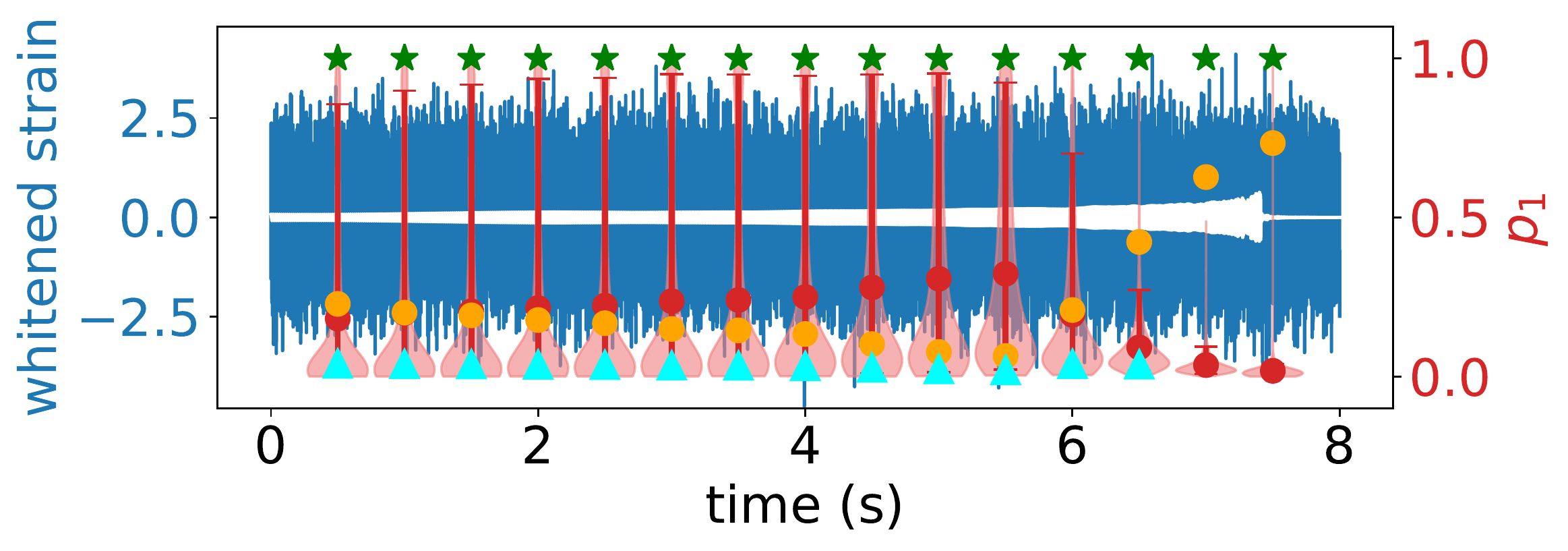}%
	\caption{\label{fig:loudBNS}The flagging result of a loud BNS event with $\rho_\text{opt}=40$.} 
\end{figure}

We further investigated the sensitivity of our model for the BNS events. 
Fig.~\ref{fig:loudBNS} shows an example where the BNS signal is as loud as $\rho_\text{opt}=40$, considered as a nearby event closer to the observer. In this case our model was aware of most time windows containing the waveform and gave low $\bar{p}_1$ with significant confidence scores near the coalescence event. 
This means that our model that was trained with the BBH events (of total component mass larger than 10 $M_\odot$) is insensitive to the BNS events.
This is not too surprising because the BNS events involve a much smaller mass range and could have quite different features in their GW waveforms as compared with the BBH events.
Therefore to possess sensitivity for the BNS events, we would need to add the BNS waveform templates into our training dataset so that our model could recognize it as a separate BNS class.
Although this process is straightforward in our framework (see Sec.~\ref{Sec:multi-class}), we did not include the BNS training in this demonstrative work due to its large variation in waveform templates thus requiring much more computation power than we had.

In summary, our model is capable of detecting the BBH events, unaffected by the treatment involving the Tukey windowing. Our current model trained with only the BBH events will not falsely detect the BNS events but could potentially raise awareness for them. In the near future a multi-class classifier will be needed for the forth-coming new GW data. We have more discussions on this in Sec.~\ref{Sec:multi-class}.

\section{\label{discussion}Discussion}

In this section we discuss various critical issues in our model, further demonstrating its strengths and future potentials.

\subsection{Extensibility for Multi-Detector Network}
\label{multi-network-detection}
In contemporary GW searches, multi-detector network is becoming crucial so the coincidence test \cite{Usman_2016,Sachdev:2019vvd, Cannon:2011vi,Adams:2015ulm,Klimenko_2008} has become critical for declaring detections. The coincidence test is based on the fact that all detectors should have observed the same event at the same time so that the triggers due to artifacts can be minimized by cross-correlating the data from different detectors. Our model is readily extensible for this as described below.

To make a neural network model capable of performing coincidence test, we need to find a way to feed the multi-detector data into the model so that the convolutional layers can perform the cross-correlation to extract useful information from the data \cite{NIPS1987_20}. Fortunately our model can readily do that by simply stacking the multi-detector data to form a multi-channel array as its input. Although this process is rather straightforward for our current setup, the size of training dataset will grow in proportion to the number of detectors, thus requiring proportionally more computation power. 

We also note that a potential advantage of the coincidence test based on the multi-detector data is the management of noise glitches. For data from a signal detector we normally need more thorough feeding of experimental glitches into the model so that our model could learn to ignore the glitches when trying to detect signals. This requires much more computation power than what we have performed. On the other hand a model including the coincidence test for multi-detector data should be able to serve the same, without the need for the model to learn all the characteristics of possible noise. We plan to demonstrate the coincidence test in our future works.

\subsection{Size of Monte-Carlo Sampling}
\label{size-of-MC-sampling}
The first is about the trade-off between efficiency and accuracy in our Bayesian Neural Network.
In theory the accuracy of BNN increases monotonically with the size of Monte-Carlo sampling, which is the main determinant of the required computation time. 
Thus there is no natural way for achieving both high accuracy and high efficiency.
For example a smaller sampling size helps speed up the prediction process while on the other hand inducing larger uncertainty in the predicted $\bar{p}_1$ and $c_1$.
To find a proper trade-off we quantified the dependence of both the prediction uncertainty and the computation time separately on the MC sampling size and also on the batch size for the flipout estimator, which enables parallel predictions (see Sec.~\ref{BNN}).

\begin{figure}[!htb]
	\centering
	\subfloat{\includegraphics[width=\linewidth]{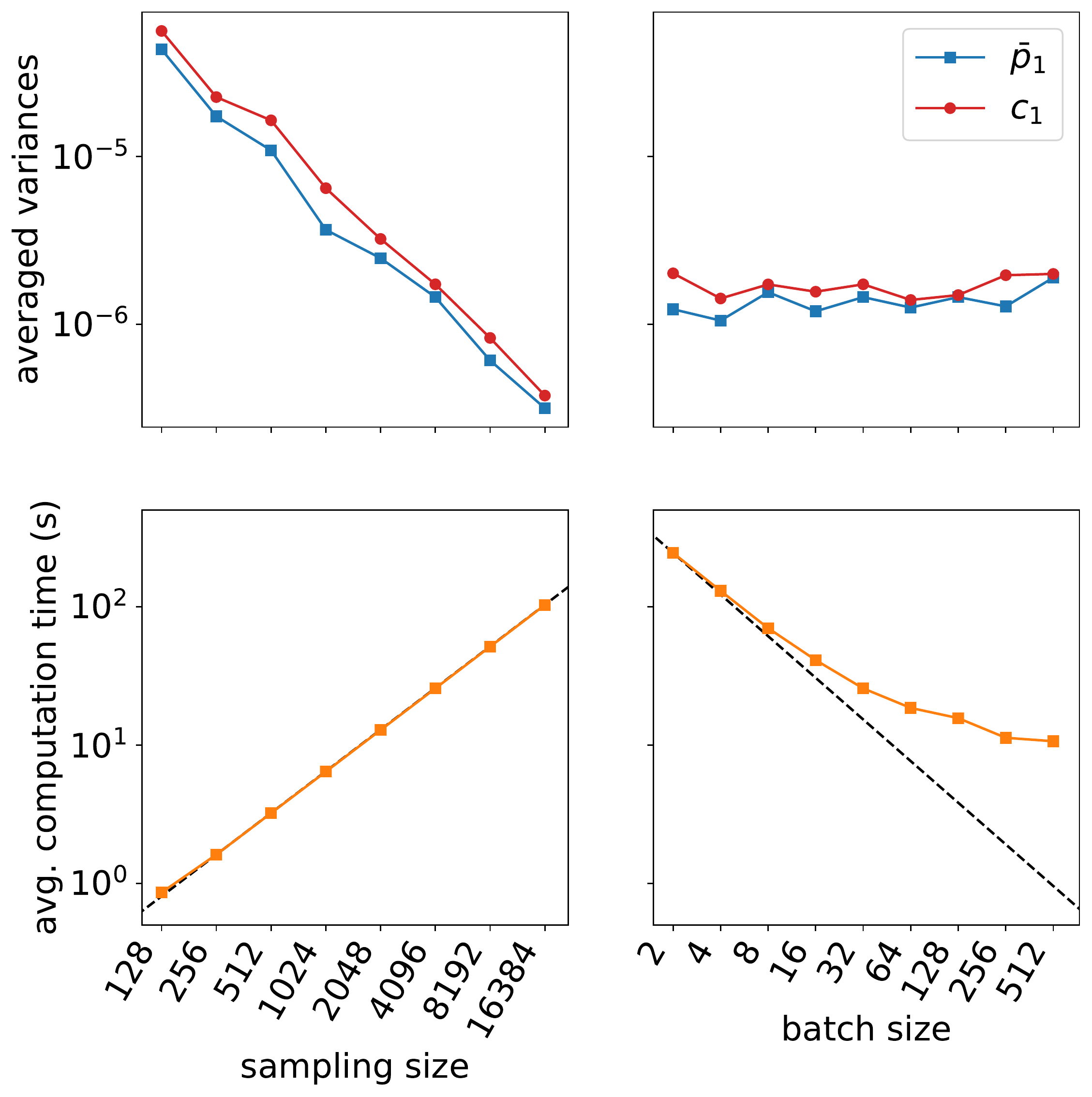}}
	\caption{\label{fig:mc_batch_test}
		The left panels show the dependence of averaged variances in $\bar{p}_1$ and $c_1$ (top) and of the average computation time (bottom) on the MC sampling size.
		The black dashed line in the bottom-left panel represents the linearity of slope $1$.
		The right panels show the dependence of averaged variances in $\bar{p}_1$ and $c_1$ (top) and of the average computation time (bottom) on the batch size for the flipout estimator, all with an MC sampling size of 4096.
		The black dashed line in the bottom-right panel represents the inverse linearity of slope $-1$.
		All results are based on the data presented in Fig.~\ref{fig:sample}
	} 
\end{figure}

About MC sampling, we considered the sizes of 128, 256, 512, 1024, 2048, 4096, 8192, and 16384, each conducted for 100 times with the data in Fig.~\ref{fig:sample} to obtain the averaged variances in $\bar{p}_1$ and $c_1$ and the average computation time.
The batch size for the flipout estimator is 32 throughout this test.
The left panels in Fig.~\ref{fig:mc_batch_test} show the results.
As expected, while both the variances in $\bar{p}_1$ and $c_1$ decrease with the sampling size, the computation time increases linearly.
Based on these results, we chose 4096 as a reasonable trade-off size for the MC sampling.

About the batch size for the flipout estimator, we considered the sizes of 2, 4, 8, 16, 32, 64, 128, 256, and 512, all with the same total sampling size of 4096 and conducted for 100 times using the data in Fig.~\ref{fig:sample}.
This test is to understand whether or not a larger batch size does improve much on the computation time, because a larger batch size actually comes with a possible price of sampling bias.
The right panels in Fig.~\ref{fig:mc_batch_test} show the results.
It is clear that the gain in computation efficiency follows the expected behavior of linearity but only for batch sizes up to about 32 but then becomes marginal for larger batch sizes.

For the above reasons, we chose to perform all the predictions in this work (as in Sec.~\ref{result}) with the sampling size of 4096 and the batch size of 32. In principle, one should increase the sampling size while keeping the batch size small whenever practically feasible.

\subsection{Model Calibration and Reliability}
\begin{figure}[!tb]
\centering
\includegraphics[width=0.98\linewidth]{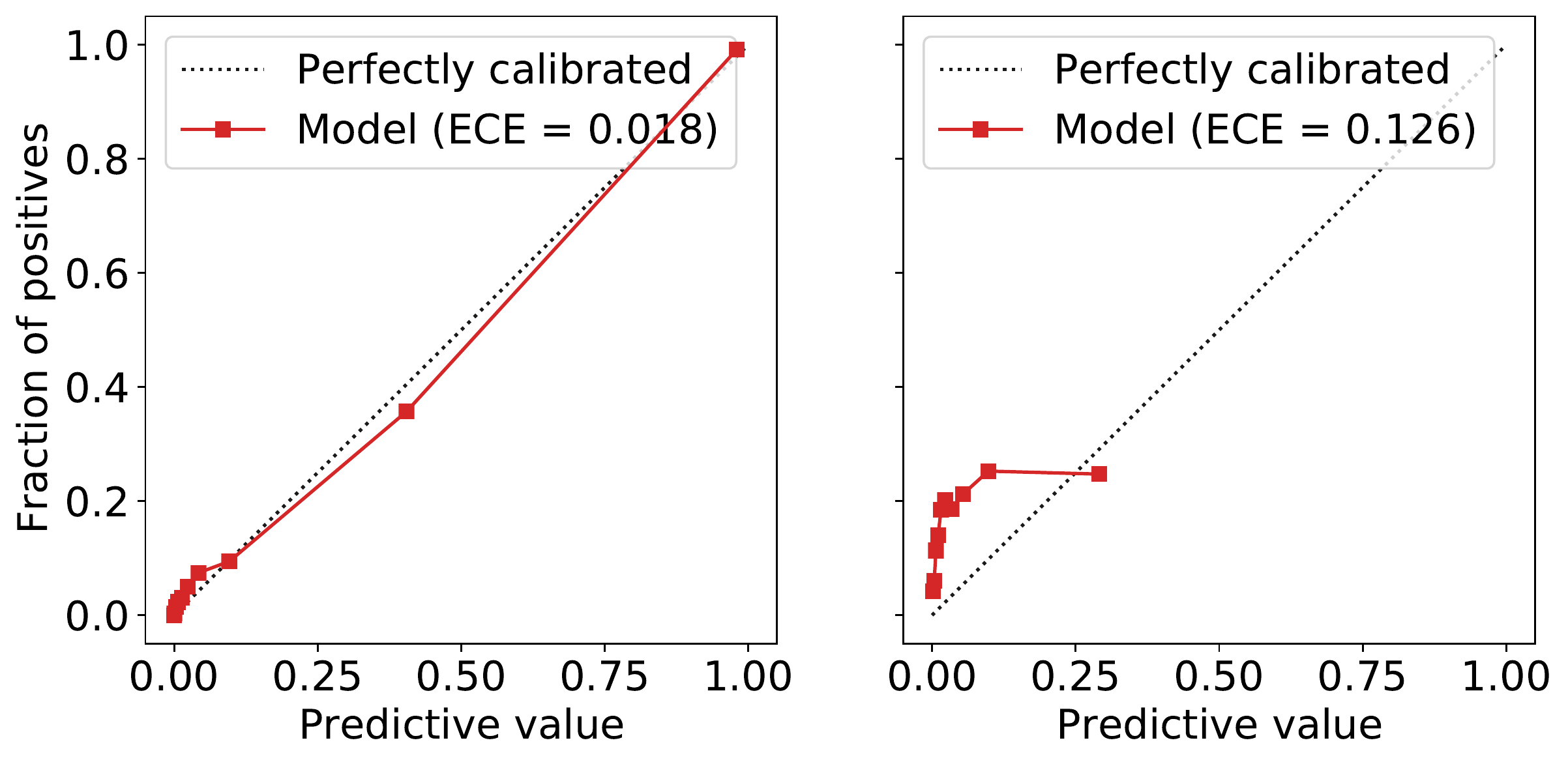}%
\caption{\label{fig:reliability}
	The reliability curves of our model based on the validation dataset	(left) and on only a subset with a low SNR of $\rho_\text{opt}=2.5$ (right).} 
\end{figure}

It has been suggested that the confidence of the predictions from a neural network classifier may be justified directly via its predictive values. 
That is to compare its predictive values with the empirical frequencies of the input data \cite{NiculescuMizil2005PredictingGP,Naeini2015,DBLP:journals/corr/GuoPSW17}. 
If they match well, this neural network classifier is said to have been well-calibrated and thus possesses confidence in its reliability.
To this end, we first followed the method in Ref.~\cite{NiculescuMizil2005PredictingGP} to produce the reliability curve, and then computed the expected calibration error (ECE) as defined in Ref.~\cite{Naeini2015}, using our validation dataset generated in Sec.~\ref{data-generation}.
Each of the 4096 samples in our validation dataset possesses 15 time windows, each with a predictive value $\bar{p}_1$. To construct the reliability curve, all these 4096$\times$15 time windows were sorted in their associated $\bar{p}_1$ and then divided into 10 bins of roughly equal size. 
Within each bin, the $\bar{p}_1$'s were averaged to become the horizontal coordinate in the left panel of Fig.~\ref{fig:reliability}, while the actual fraction of real events (the positives) formed the vertical coordinate.
The 10 bins thus generated 10 points in the plot, linked to form the reliability curve (the red solid line). Ideally we would like this curve to be as close as possible to the perfect curve, which has a slope of one through the origin (the dotted line).
It is clear that our model is fairly well-calibrated with a very small error of $\text{ECE}=0.018$. 
We could thus conclude that the $\bar{p}_1$ generated by our model is a confident estimate for reality.

However, such confidence may break down in the following two cases. The first is when the input data does not follow the same distribution as the training dataset. The fact that our model, currently trained with only the BBH events, is insensitive to the BNS events is one example (see Sec.\ref{performance-against-real-data}).
This is why we introduced and utilized the confidence score $c_k$ (see Eq.~(\ref{eq:conf_score})) in our work. The confidence score would be low whenever the input data do not comply with the underlying distribution of the training data.
This out-of-distribution problem can be minimized by training our model with datasets of all possible kinds that we may encounter during the observation.

The second case for the confidence breakdown is when the signal-to-noise radio is not high enough so that the model fails to detect all the features it is supposed to recognize.
To demonstrate this point, we computed the reliability curve for a subset of our validation data with $\rho_\text{opt}=2.5$.
The right panel of Fig.~\ref{fig:reliability} shows the result. The reliability curve (red solid) being well to the left of the perfect curve (dotted) with a significant error of $\text{ECE}=0.126$ indicates under detection of real events (the positives). This is simply due to the obscuration of the GW signals by the noise. Such a problem can be readily improved by deepening our model training to gain higher sensitivity.

To sum up, we may train our model deeper to gain higher sensitivity and also feed it with more signal types to avoid the breakdown of confidence. Furthermore, as long as our model possesses high enough sensitivity through deep training, our uncertainty estimation based on the confidence score could serve as a powerful tool for picking up the out-of-distribution signals that we have never seen before.

\subsection{Multi-Class Detection}
\label{Sec:multi-class}
In this work we only trained our model with BBH signals to perform binary classification. However, as more BNS signals have been discovered by LIGO and Virgo \cite{TheLIGOScientific:2017qsa,Abbott:2020uma}, a classifier that can further distinguish among different types of signals such as those from BBH, BNS and BHNS is needed. For example the deterministic CNN proposed in Ref.~\cite{Krastev:2019koe} is able to distinguish between the BBH and BNS signals. 
A multi-class classifier can not only reduce the search space of matched-filtering but also improve the uncertainty estimation in the Bayesian approach that we propose here.
To estimate the uncertainties in signal predictions for a multi-class Bayesian approach, we need to employ the covariance matrix defined in Equation \eqref{eq:uncertainty_matrix}, which is more complicated than Equation \eqref{eq:kthuncertainty} used in this work. In turn the definition of the confidence score in Equation \eqref{eq:conf_score} also needs to be updated accordingly.
We will leave these to future works.

\subsection{Stride Size of Sliding Window}

\begin{figure*}[!htb]
	\thisfloatpagestyle{empty}
	\centering
	\subfloat{\includegraphics[clip,width=\linewidth]{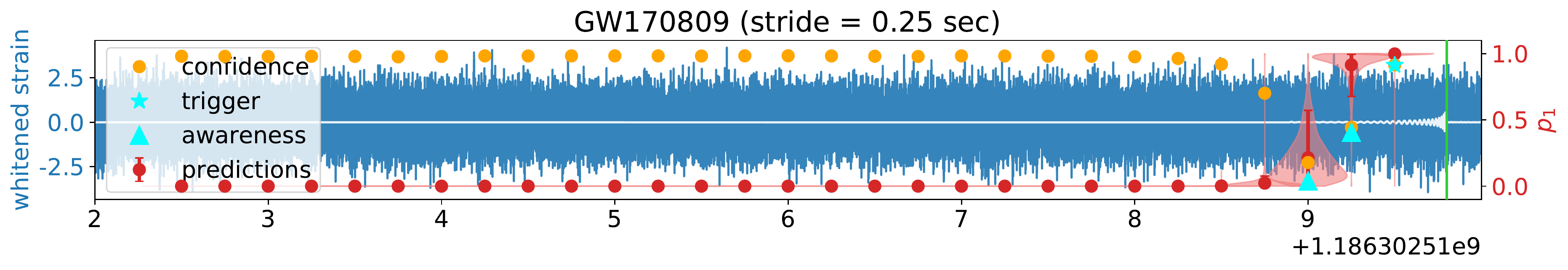}}\\[-3.12pt]
	\subfloat{\includegraphics[clip,width=\linewidth]{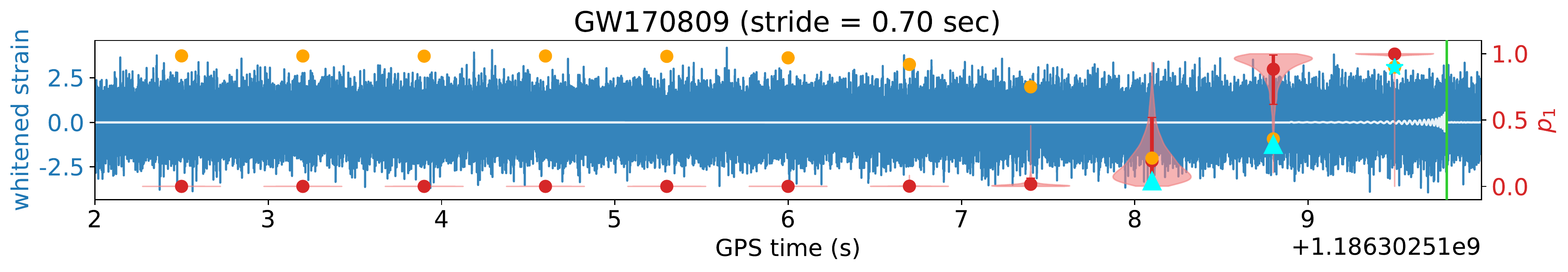}}\\[-3.12pt]
	\caption{\label{fig:LIGO_170809_stride}The flagging results of the Livingston GW170809 data using sliding-window search with stride sizes of 0.25 second (top) and 0.7 second (bottom).}
\end{figure*}

The architecture of our model combines the CNN model and the sliding-window searches. Here we illustrate the effect of the stride size of our sliding windows.
In principle, while keeping the same window size, a smaller stride size will increase the time resolution in pinning down a GW event.
Fig.~\ref{fig:LIGO_170809_stride} shows the flagging results of the LIGO Livingston GW170809 data using stride sizes of 0.25 (top) and 0.7 (bottom) second, to be compared with our previous result for using 0.5 second (the fourth panel in Fig.~\ref{fig:LIGO_BBH}).
All time windows have the same size of one second.
It is evident that the 0.25-second result demonstrates the best accuracy in labeling the temporal location of the GW event.

The only drawback for a smaller stride size is that the computation time is inversely increasing with the stride size, so there is always a trade-off given the available computation power.
Another issue concerning the stride size is that a stride size larger than half of the window size will cause an un-uniform coverage of the time domain in the flagging analysis, leading to possible misses for the GW events \cite{Gebhard:2019ldz}.
Therefore it is optimal to choose a stride size of half the window size in face of the always limited computation power. In the main work presented in this paper we have chosen a window size of one second with a stride size of 0.5 second.
When analyzing the forthcoming new data in future, one can always use our model here to first efficiently identify a GW event, hopefully in nearly real time, and then focus on the identified data chunks with increased time resolution and deeper search or even further using the matched-filtering search to pinpoint the coalescence time.

\subsection{Forecasting GW Events}

It is always desirable to be able to detect the GW waves before a coalescence event so that we could conduct parallel targeted observations such as the EM observations to trace the full process including the inspiral, merger and ringdown phases.
A similar desire in the trading market has actually led to using the RNN model to perform time-series forecasting in detecting the anomaly in time series \cite{malhorta2015} and in predicting the stock price \cite{2017arXiv171005751E}.
However our techniques based on deep learning for detecting GW waves rely strongly on recognizing the features of the GW waveform around the coalescence time, where the SNR is much stronger. This in turn implies that our model could detect the GW signals likely only when a coalescence event has come into the scene, and then we trace back in time to identify the data sections that may have contained the GW waveform before the event. 

\begin{figure}[!htb]
	\thisfloatpagestyle{empty}
	\centering
	\subfloat{\includegraphics[clip,width=\linewidth]{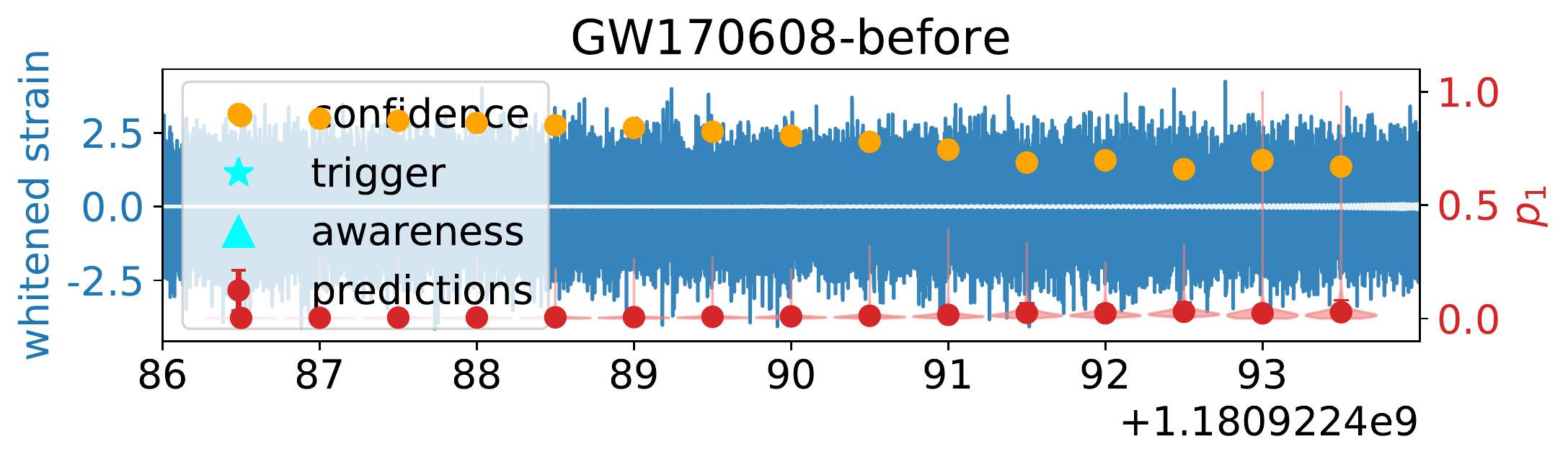}}\\[-3.12pt]
	\subfloat{\includegraphics[clip,width=\linewidth]{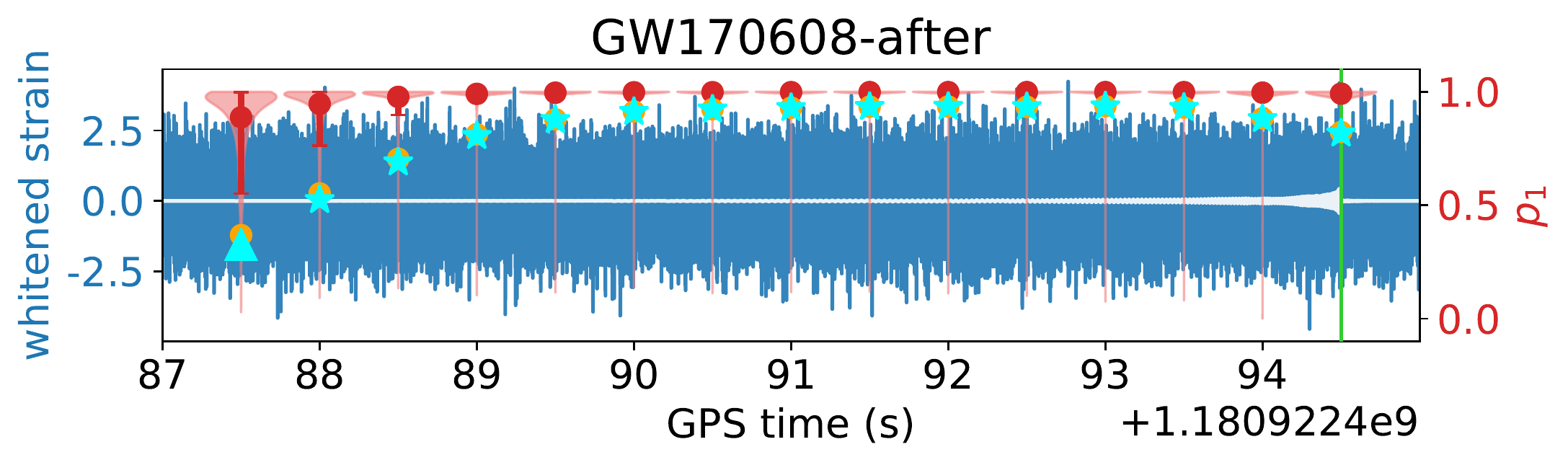}}\\[-3.12pt]
	\caption{\label{fig:LIGO_170608}
		The flagging results on the Livingston GW170608 data without the coalescence event (top) and with the event (bottom). The only difference between these two sub-datasets is the inclusion of the extra one second (on the right) in the bottom panel where the event sits in the middle of this second.}
\end{figure}

We verify this here using the Livingston GW170608 data.
The results are shown in Fig.~\ref{fig:LIGO_170608}, where the bottom panel contains only one extra second where the coalescence event sits in its middle.
The top panel shows no sign of GW detection even though the event is about to enter our time domain of analysis, only half a second away.
Once the event enters our time domain (bottom panel), our model immediately picks up all the time windows containing not only the event but also the waveform prior to the event.
This is not surprising because for each forward pass our model uses the zero states as the initial hidden and cell states. To enable the capability for prior detections, we have to incorporate the hidden and cell states of the previous prediction into our model. This would require not only modifications in our model structure but also the construction of partly correlated samples in the training dataset.
An extra complication is that the Bayesian Neural Network would need to have all the previous hidden and cell states in the form of distributions. Considering these challenges we will leave the attempts for forecasting to future work.

\section{\label{conclusion}Conclusion}
We have proposed a new model to demonstrate the capability of Bayesian Neural Network in detecting gravitational waves.
One critical bonus for using the BNN is the ability for uncertainty estimation, which is particularly useful when facing the data that do not follow the distribution of the training dataset. Our uncertainty estimation is manifested by the newly defined confidence score $c_1$, which in turn defines the `awareness state' for collecting the cases where triggers cannot be accepted nor rejected with confidence.
These cases can then be further investigated in the follow-up checks.
In addition, our Bayesian CLDNN model integrates the CNN classifier with the sliding-window search scheme so that we could detect most of the time windows that contain the GW waveform in a coalescence event.
However due to the limited computation power of this work, our current model does not outperform the sensitivity of the existing matched-filtering search but still successfully detect all the GW events that LIGO detected in the O2 observation. Nevertheless on a moderate GPU-equipped personal computer, it takes only about 20 seconds for detecting an event and labeling its waveform period whenever a coalescence event comes into the time domain of our analysis. 
This 20-second latency is expected to be dramatically improved by a GPU-optimized code with enhanced computation power and even shorter data chunks, making our model possible for nearly real-time detection. 
Such nearly real-time detection is unlikely to be possible for the matched-filtering search, which is always computationally expensive.
In future we plan to further explore the discussed potential for a Bayesian CLDNN model in forecasting the GW coalescence events.

\begin{acknowledgments}
We would like to thank Chun Hsiang Chan for useful discussions on deep learning and statistics. This research has used the data, software and/or web tools from the Gravitational Wave Open Science Center (https://www.gw-openscience.org), a service of LIGO Laboratory, the LIGO Scientific Collaboration and the Virgo Collaboration. LIGO is funded by the U.S.~National Science Foundation. Virgo is funded by the French Centre National de Recherche Scientifique (CNRS), the Italian Istituto Nazionale della Fisica Nucleare (INFN) and the Dutch Nikhef, with contributions by Polish and Hungarian institutes.
We acknowledge the support from Ministry of Science and Technology, Taiwan.
\end{acknowledgments}%

\bibliography{Draft}

\end{document}